\documentclass[usenatbib]{mnras}

\usepackage[utf8]{inputenc}
\usepackage{graphicx}
\usepackage{amsmath}
\usepackage{siunitx}
\usepackage{placeins}
\usepackage{comment,xcolor}
\usepackage{appendix}
\usepackage{caption}
\usepackage{subcaption}
\usepackage{gensymb}
\usepackage{float}
\usepackage{ulem}
\usepackage{xcolor}

\definecolor{dg}{rgb}{0.0, 0.6, 0.1}

\newcommand{\bfm}[1]{\mbox{\boldmath$ #1 $}}
\graphicspath{{Pictures/}}
\title[A Galactic breeze and the Fermi bubbles]{A Galactic breeze origin for the Fermi bubbles emission}

\author[O.~Tourmente et al.]{
Olivier~Tourmente$^{1}$\thanks{E-mail: olivier.asin@desy.de},
D.~Rodgers-Lee$^{2,3}$ and
Andrew~M.~Taylor$^{1}$
\\
$^{1}$Deutsches Elektronen-Synchrotron, Zeuthen, Germany\\
$^{2}$School of Cosmic Physics, Dublin Institute for Advanced Studies, 31 Fitzwilliam Place, D02 XF86, Ireland \\
$^{3}$School of Physics, Trinity College Dublin, University of Dublin, College Green, Dublin 2, D02 PN40, Ireland 
}
\date{Accepted XXX. Received YYY; in original form ZZZ}
\pubyear{2022} % Enter the current year, for the copyright statements etc.

\begin{document}
\maketitle

\begin{abstract}

The origin of the Fermi bubbles, which constitute two gamma-ray emitting lobes above and below the Galactic plane, remains unclear. The possibility that this Fermi bubbles gamma-ray emission originates from hadronic cosmic rays advected by a subsonic Galactic outflow, or breeze, is here explored. The simulation of a breeze solution and subsequent cosmic ray transport is carried out using the hydrodynamical code, PLUTO, in combination with a cosmic ray transport code. The Galactic outflow model obtained is found to be compatible with both inferences of the decelerating outflow velocity profile of the gas in the Fermi bubbles region, and evidence for the presence of a large amount of hot ionised gas out in the Galactic halo region. 
Although simple, this model is found to be able to reproduce the observed Fermi-LAT energy flux at high Galactic latitudes. Following these results a prediction concerning the gamma-ray emission for 1-3~TeV photons is made for future comparison with CTA/SWGO measurements. 

\end{abstract}


\section{Introduction}

Observations and measurements have provided evidence of outflows in galactic centers (GCs) for several decades \citep{1963ApJ...137.1005L, 1964ApJ...140..942B, 1971ApJ...170..241M, 1997ApJ...490L...1S}. Such outflows have also been observed from the Milky-Way in a broad energy range from radio \citep{1984ApJ...283...90L} to X-ray \citep{1996ARA&A..34..645M, 1997ApJ...481L..43C}.

A decade ago, two Galactic bubbles lobes extending above and below the GC were first detected with the Fermi-LAT instrument in the gamma-ray energy band \citep{2010, Dobler_2010}. The extension of this emission is $\sim$50$^{\circ}$ in Galactic latitude and $\sim$40$^{\circ}$ in Galactic longitude. Furthermore, the spectral energy distribution of this gamma-ray emission is rather hard, exhibiting a spectral downturn or cutoff at an energy of $\sim 300$~GeV \citep{Ackermann_2014}. The brightness intensity of this emission appears approximately constant within the bubbles, with a sudden sharp edge feature observed at their boundary \citep{Ackermann_2014}. Those gamma-ray bubbles are potentially a counterpart of the microwave haze \citep{2004ApJ...614..186F, 2013A&A...554A.139P}, with both exhibiting hard spectra. In the X-ray band, recent observations of larger scale eROSITA bubbles, which appear to enshroud the Fermi bubbles, are likely connected to the Fermi bubbles structures \citep{2020}. 

Despite the wealth of observational data, there remains no clear consensus about the underlying mechanism that produces the outflow responsible for the formation of those bubbles. Three features must be fully explained to give a satisfying theory \citep{2018}. These are related to the emission mechanism, the location of the event and the acceleration mechanism of the cosmic rays (CRs) that are responsible for the gamma-ray emission. Considering the hard spectrum of the Fermi bubbles, the emission may be generated by hadronic CR energy losses. In such scenarios, hadronic CRs are transported via starburst or active galactic nucleus winds \citep{2011PhRvL.106j1102C, 2014ApJ...790..109M, 2015ApJ...808..107C, 2015ApJ...811...37M}. The wind velocities in these models range from a few hundred of km~s$^{-1}$ to a few thousand of km~s$^{-1}$ and the timescale for the bubbles to be formed is between $\sim$10~Myr to 1~Gyr. 

Alternatively, the Fermi bubbles emission could also be produced by leptonic CR energy losses, with such leptons first being transported via an active galactic nucleus jet  \citep{2012ApJ...756..181G, 2012ApJ...761..185Y, 2013MNRAS.436.2734Y, 2017ApJ...850....2Y, 2017IAUS..322..189G}. For these models, the short cooling time of electrons implies that the age of the Fermi bubbles should be no more than a few Myr. Alternatively, these CR leptons may be accelerated near the gamma-ray production site \citep{2011ApJ...731L..17C, 2017MNRAS.467.3544S, 2019A&A...622A.203M}.    

Even if the hadronic wind models exhibit a much lower velocity than the leptonic jet models the majority of proposed models assume a supersonic velocity above $\sim$ 500~km~s$^{-1}$. However, UV absorption line observations of cold clouds in the Fermi bubbles, from low to high Galactic latitudes, show velocities ranging from $\sim$ 100 - 300~km~s$^{-1}$ and with a continuous deceleration  \citep{Fox_2015, 2017ApJ...834..191B, 2018ApJ...860...98K, 2020ApJ...888...51L, 2020ApJ...898..128A, 2021ApJ...923L..11C, 2022PASJ..tmp...58S}. This last feature contradicts the supersonic wind model for which a shock should be seen producing strong perturbations in the velocity profile and in the ambient density. 

Small Galactic outflow velocities are consistent with expectations for a thermally driven outflow \citep{1958ApJ...128..664P, 1965ApJ...141.1463P, 1965ApJ...141..320C}. This model is motivated from considerations of the momentum transport equation, which provides two possible solutions, a transonic and a subsonic one. The transonic description has previously been applied in galactic outflow models \citep{1970ARA&A...8...31H, 1985Natur.317...44C, 2007ApJ...656...93E} and galactic fountain models \citep{Avillez2004, Bustard2018, Chan2021}. The subsonic description is less explored, with only stability analysis of such outflows having been made \citep{Fichtner1997}. 

Following the first UV absorption line observations of cold clouds expanding in the Fermi bubbles, \citet{2017PhRvD..95b3001T} proposed an alternative where CRs are both diffused and advected within a Galactic breeze. The subsonic profile that \citet{2017PhRvD..95b3001T} used was expressed as an analytic, divergence-free outflow velocity profile. This outflow with pre-accelerated hadronic CRs at its base was launched from the GC, subsequently producing $\gamma$-ray emission via proton-proton interactions with the diffuse ambient gas out in the halo region. In this work, the model is investigated further by relaxing the divergence-free assumption for the subsonic outflow by using a hydrodynamics code. The velocity profile thus produced is used in the CR transport code, taking into account advective processes. The model is a first step to study the feasibility of a Galactic breeze solution for reproducing the Fermi bubbles emission. Therefore, a single isothermal Galactic gas temperature has been assumed and the model does not include a magnetic field but instead adopts a single homogeneous diffusion coefficient. 

The paper is structured as follows: in section \ref{Galactic_model} the Galactic hydrodynamical model and the gravitational potential are introduced, as well as the numerical setup. Section \ref{CRs_model} described the transport equation for hadronic CRs and the computational grid. The subsequent results are presented in section \ref{Results} and the conclusions and outlook are given in section \ref{Conclusions}.

\section{Galactic Outflow Model}\label{Galactic_model} 

To simulate the launching region and propagation of an outflow into the Galactic halo region, a hydrodynamic description has been simulated using the PLUTO code \citep{2007}. This numerical scheme solves the mass and momentum conservation equations throughout the computational grid domain,
\begin{align}
     \frac{\partial \rho}{\partial t} + \nabla\cdot(\rho \textbf{v}) &= S_{\rho} \label{Mass_conservation},\\
     \frac{\partial}{\partial t}(\rho \textbf{v}) + \nabla\cdot(\rho \textbf{v}\textbf{v} + p\bfm{I}) &= -\rho \nabla \Phi_{\rm eff}. \label{Momentun_conservation}
\end{align}

\begin{comment}
     \frac{\partial}{\partial t}(\rho e + \rho \Phi) +\nabla\cdot\left[(\rho e + \rho \Phi + p)\textbf{v}\right] &= S_e
\end{comment}

In these expressions, $\rho$, $\textbf{v}$ and $p$ are the mass density, velocity and thermal pressure of the gas in the Galaxy, respectively. $\bf{I}$ is the unitary tensor. The term $S_{\rho}$ represents the injected mass and $\Phi_{\rm{eff}}$ is the effective gravitational potential (see section \ref{effective}) determined by first calculating the total gravitational potential, $\Phi$. Since an isothermal gas temperature is adopted (see section \ref{grid}), here the energy conservation equation has been omitted. Details about the computational grid are provided in section \ref{grid}.
It should be noted that the effect of the presence of CRs is not included in the hydrodynamical simulation, with the CRs assumed to have no dynamical effect. Following our simulation calculations, for all simulation time steps, the thermal pressure is found to be superior to the CR pressure throughout the simulation grid, except at the injection site where the CR pressure is higher by a factor of $\sim 1.8$.

\subsection{Galactic gravitational potential}\label{potential}

\begin{figure}
    \centering
    \includegraphics[scale=0.6]{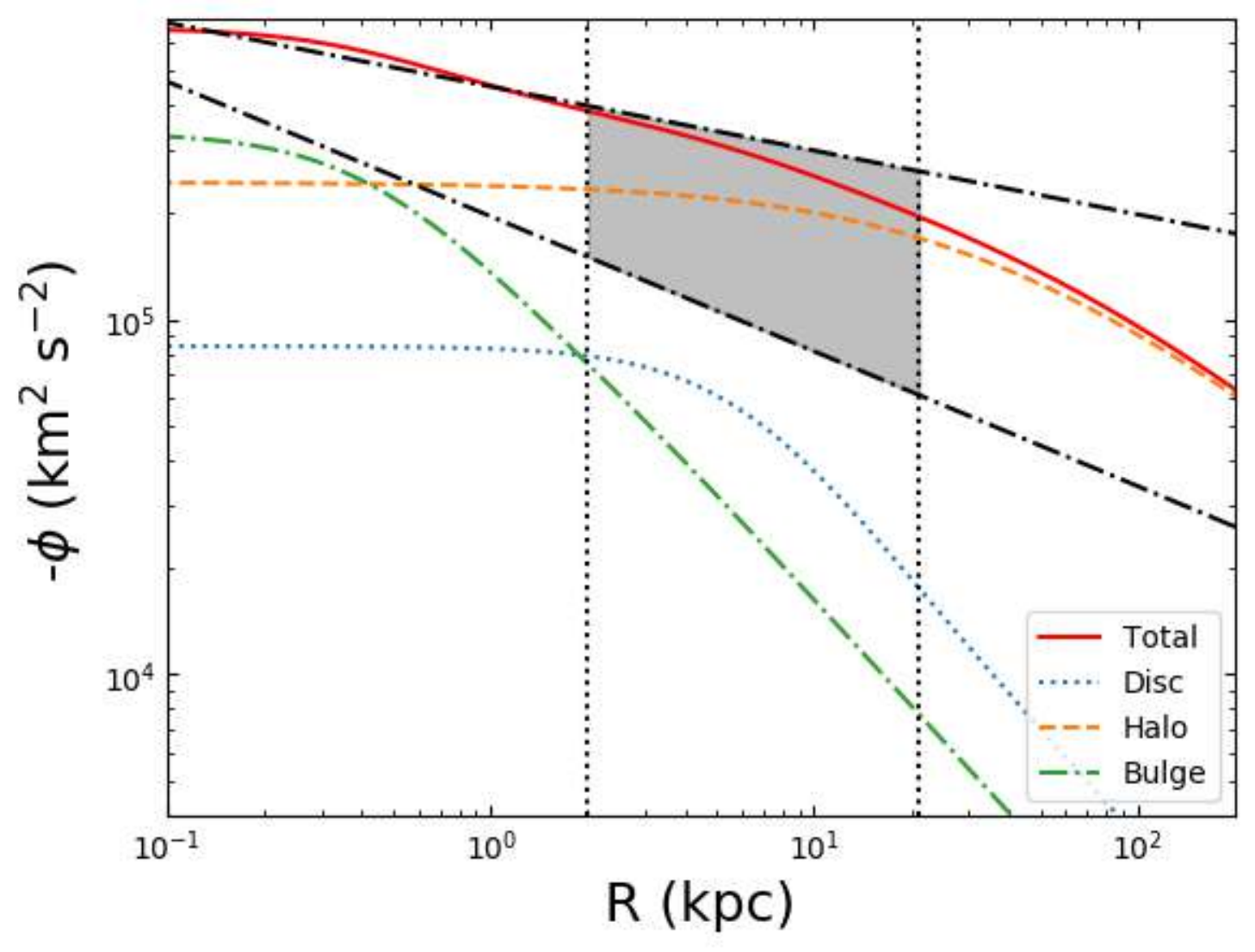}
    \caption{Determination of the total gravitational potential of the Milky Way (solid red line) based on observations from the Gaia mission \citep{2019ApJ...873..118W}. The total potential is composed of a bulge (dot-dashed green line), disc (dotted blue line) and dark matter halo component (dashed orange line). The two vertical dotted black lines at $R=$ 2 and 21~kpc indicate the spatial range for which the inferred Gaia constraint is applicable. The dot-dashed black lines show the extent of the best-fitting power laws from the Gaia observations in the 2-21~kpc region and their extension beyond this range. } 
    \label{fit_potential}
\end{figure}

The total Galactic gravitational potential, $\Phi$, results from the sum of three components, a bulge ($\Phi_{\rm b}$), disc ($\Phi_{\rm d}$), and dark matter halo ($\Phi_{\rm DM}$) such that
\begin{equation}
    \Phi = \Phi_{\rm b} + \Phi_{\rm d} + \Phi_{\rm DM}.
\end{equation}
The Gaia mission \citep{2016A&A...595A...1G, 2016A&A...595A...2G, 2018A&A...616A...1G} has provided proper motions for approximately one billion stars. With these new data a range of power-laws have been provided for fitting the Milky Way's gravitational potential \citep{2019ApJ...873..118W} giving a constraint for simulations. In figure \ref{fit_potential} those fitting power-laws (dot-dashed black lines) have been compared with $\Phi$ (solid red line) used for the simulations here. Next, we describe each component of the gravitational potential.

\subsubsection{The bulge potential}

The central bulge potential (dot-dashed green line in figure \ref{fit_potential}), $\Phi_{\rm b}$,  is based on \citet{1996} and is described by
\begin{equation}
    \Phi_{\rm b} = \sum^2_{i=1}-\frac{GM_{\rm C_{\rm i}}}{\sqrt{r^2 + r^2_{\rm C_{\rm i}}}},
\end{equation}

\noindent where $r$ is the spherical radius and $M_{\rm C_{\rm 1}}$, $M_{\rm C_{\rm 2}}$, $r_{\rm C_{\rm 1}}$ and $r_{\rm C_{\rm 2}}$ are given in table \ref{table:1}.

\subsubsection{The disc potential}

The disc potential (dotted blue line in figure \ref{fit_potential}), $\Phi_{\rm d}$, is based on \citet{1996} which is a three component Miyamoto-Nagai potential \citep{1975PASJ...27..533M} given by
\begin{equation}
    \Phi_{\rm d} = \sum^3_{i=1}-\frac{GM_{D_{\rm i}}}{\sqrt{R^2 + \left(a_{\rm i} + \sqrt{z^2 + b^2}\right)^2}},
\end{equation}
where  $R$ and $z$ are the cylindrical coordinates and $M_{\rm D_{1}}$, $M_{\rm D_{2}}$, $M_{\rm D_{3}}$, $a_1$, $a_2$, $a_3$ and $b$ are given in table \ref{table:1}.

\subsubsection{The dark matter halo potential}

The dark matter halo potential (dashed orange line in figure \ref{fit_potential}), $\Phi_{\rm DM}$, is given by a Navarro-Frenk-White (NFW) profile \citep{1996_NFW},
\begin{equation}
    \Phi_{\rm DM} = -\frac{GM_{200}}{r_sf(c_{200})}\frac{\ln{(1+r/r_s)}}{r/r_s},
\end{equation}
where the function $f(c_{200}) = \ln{(1 + c_{200})} - c_{200}/(1 + c_{200})$, $M_{200} = 2\times10^{12}\ M_{\odot}$ \citep{2016MNRAS.461.3483T} and $c_{200}$ is the concentration parameter defined as $c_{200} = r_{200}/r_s$ where $r_{200} = 258$~kpc and $r_s = 21.5$~kpc is the scale radius. 

\subsubsection{The effective Galactic potential}\label{effective}
Observations of the Milky Way's rotation curve within the disc indicate velocities of $\sim$200~km~s$^{-1}$ beyond 2~kpc from the GC \citep{2013PASJ...65..118S,2014ApJ...785...63B}. Eqs.~(\ref{Mass_conservation}) and (\ref{Momentun_conservation}) have then been solved in a rotating frame since such a Galactic rotational velocity value cannot be neglected for a Galactic breeze model, for which the outflow velocity is similar in magnitude to the Galactic rotation velocity. In the rotating frame, the effective Galactic potential, $\Phi_{\rm eff}$, used in Eq.~(\ref{Momentun_conservation}) is of the form,
\begin{equation}
    \Phi_{\rm eff} = \Phi + \frac{R}{2}\left(\frac{d\Phi}{dR}\right).
\end{equation}

\begin{table}
\begin{center}
\begin{tabular}{c c c}
\hline
Component & Parameter & Value \\
\hline
 $\Phi_{\rm b}$ & $r_{C_{1}}$ & 2.7 kpc \\
 & $M_{\rm C_{1}}$ & 6.0$\times 10^{9}\ M_{\odot}$ \\
 & $r_{\rm C_{2}}$ & 0.42 kpc \\
 & $M_{\rm C_{2}}$ & 3.2$\times 10^{10}\ M_{\odot}$ \\
 \\
$\Phi_{\rm d}$ & $b$ & 0.3 kpc \\
 &$a_1$ & 5.81 kpc \\
 & $M_{\rm D_{1}}$ & 1.32$\times 10^{11}\ M_{\odot}$ \\
 & $a_2$ & 17.43 kpc \\
 & $M_{\rm D_{2}}$ & -5.8$\times 10^{10}\ M_{\odot}$ \\
  & $a_3$ & 34.86 kpc \\
 & $M_{\rm D_{3}}$ & 6.6$\times 10^{9}\ M_{\odot}$ \\
 \\
 $\Phi_{\rm DM}$ & $r_s$  & 21.5 kpc \\
\hline
\end{tabular}
\caption{List of parameters for the Galactic gravitational potential model.}
\label{table:1}
\end{center}
\end{table}

%\subsubsection{Fitting of the Milky Way potential}

\subsection{1D spherical Galactic outflow}
\label{1D_outflow}

Some intuition on the outflow evolution can be found through the consideration of a 1D spherically symmetric
outflow \citep{1958ApJ...128..664P, 1965ApJ...141.1463P, 1965ApJ...141..320C, lamers_cassinelli_1999}. For this case, mass flux conservation leads to the relation
\begin{equation}\label{mass}
    \dot{M} = 4\pi r^2 \rho(r) v(r),
\end{equation}
where $\dot{M}$ is defined as the mass loss rate. The momentum flux conservation equation, assuming an isothermal, steady-state evolution, leads to the relation
\begin{equation}
    v\frac{dv}{dr} + \frac{c_s^2}{\rho}\frac{d\rho}{dr} + \frac{d\Phi}{dr} = 0,
\end{equation}
where $c_s$ is the thermal velocity. By differentiating Eq.~(\ref{Mass_conservation}) to get the density gradient and substituting it into Eq.~(\ref{Momentun_conservation}) the following expression is obtained 
\begin{equation}\label{outflow}
    \frac{1}{v}\frac{dv}{dr} = \frac{1}{r}\left(\frac{2c^2_s-\frac{rd\Phi}{dr}}{v^2-c^2_s}\right).
\end{equation}
The numerator on the right-hand side of this expression goes to zero when $r\frac{d\Phi}{dr} = 2c^2_s$.
The radius where this occurs is called the critical radius, $r_c$.
Beyond $r_c$,  for a positive numerator, a transonic outflow will accelerate while a subsonic outflow will decelerate. 
The behavior of a transonic outflow and its different solutions have been studied in a gravitational potential of a cold dark matter halo \citep{2013MNRAS.432.2837T, 2014MNRAS.444.1177I, 2017MNRAS.470.2225I}. Moreover a transonic outflow considerably reduces the ambient density whereas a subsonic outflow does not. These effects are described in greater detail in appendix \ref{Hydrostatic}.

The total effective potential model adopted for the Galaxy is not entirely spherically symmetric due to both the disc component and the Galactic rotation. These effects have an influence on the determination of the value of $r_c$. This value is important since it determines the subsequent velocity evolution which is discussed in the following section. 

\subsection{Galactic gas temperature and density profile}\label{grid}

As discussed at the beginning of section \ref{Galactic_model}, the hydrodynamic equations (see Eq.~(\ref{Mass_conservation}) and (\ref{Momentun_conservation})) are considered for an isothermal gas. Indeed, observations of the Galactic halo gas do not show strong temperature fluctuations \citep{2017ARA&A..55..389T}. 

Based on \citet{2019ApJ...873..118W}, the normalisation for the Galactic potential, $\Phi$, has been adopted with a value at the upper end of the allowed range (see Fig. \ref{fit_potential} and section \ref{potential}). In order to maximise the outflow velocity, while maintaining the existence of $r_c$ beyond the radius where the CRs are injected, a gas temperature of $kT \approx 400~\rm{eV}$ (ie. $c_s\approx 250~\rm{km}\,\rm{s}^{-1}$) is adopted. By choosing these values for $\Phi$ and $c_s$  implies $r_c\approx 1~\rm{kpc}$ (see section \ref{1D_outflow}). This value of $kT \approx 400~\rm{eV}$ is compatible with recent observations of the Galactic halo that presents a mixed hot-warm phase with a hot phase for which the temperature can reach $\sim$1~keV \citep{Das_2021}. It should be noted that a small increase in the gas temperature does not significantly change the critical radius and thus has little effect on the results.

For the initial gas density setup, a hydrostatic density distribution is assumed, dictated by the Galactic potential, $\Phi$, and the single gas temperature with thermal velocity, $c_{s}$. This initial spatial gas number density distribution can be expressed as,
\begin{equation}\label{density}
    n_{\rm{halo}} = n_{10\rm{kpc}}\exp{\left(-\frac{\Phi}{c^2_s}\right)},
\end{equation}

\noindent where $n_{10\rm{kpc}}$ is a normalisation constant. Its value has been fixed at $n_{10\rm{kpc}} = 10^{-3}~\rm{cm}^{-3}$ in order to be compatible with the observational values provided in a range of 10-100~kpc \citep{2017ApJ...836..243G,Martynenko_2022}. Fig.\ref{hydrostatic_density} in Appendix~\ref{Hydrostatic}  shows the hydrostatic gas number density distribution adopted in this paper.

\subsection{Numerical setup}\label{Pluto_setup}

As shown in section \ref{1D_outflow}, Eq.~(\ref{outflow}) implies that an outflow driven by thermal gas pressure can reach a given velocity at $r_c$ for a particular velocity $v_0$ at the launching zone $r_0$. For the simulations, the combination of values adopted for $v_0$ and the thermal temperature  collectively dictate the maximum velocity of the outflow at the critical radius, $r_c$. Although a subsonic solution at $r_c$ only requires the outflow to reach a Mach number $M<1$, for our simulations the outflow reaches $M=0.85$ at $r_c$. The inner radius, $r_0$, is fixed to be $r_0 = 300~\rm{pc}$. The thermal pressure at $r_0$ is then defined as $P_0 = \rho_0c_s^2$.
The computational grid has been setup as a 2D spherical grid ($r$, $\theta$). A logarithmic grid is used with 256 bins in the $r$-direction and uniform grid spacing in the $\theta$-direction with 90 bins. The outflow is launched at $r_0$ and the outer radial boundary is set to an absorbing boundary condition, at 300~kpc. With this setup, the outflow does not reach the outer boundary during the simulation time of 300~Myr. The system reaches steady state at smaller distances at earlier times than this. Both the inner and outer boundary in the $\theta$-direction have been set to reflective. The simulation domain covers a single hemisphere, with $\theta$ ranging from 0 to $\pi/2$.

\section{CR transport model}\label{CRs_model}

The transport of CRs through an outflow is described, in its conservative form, by the following equation,

\begin{equation}\label{CRs_transport}
    \frac{\partial f}{\partial t} = \bfm{\nabla}\cdot\left(D\bfm{\nabla} f - \bfm{v}f\right) + \frac{1}{p^2}\frac{\partial}{\partial p}\left[\left(\bfm{\nabla}\cdot \bfm{v}\right)\frac{p^3}{3}f\right] - \frac{f}{\tau_{\rm loss}} + \frac{Q}{p^2}
\end{equation}

where $f = \dfrac{dN}{d^{3}x d^{3}p}$ is the CR phase space density and $N$ is the number of particles. $D$ is the spatial diffusion coefficient, $v$ is the outflow velocity, $p$ is the CR momentum, $\tau_{\rm loss}$ is the energy loss time scale for $pp$ collisions and $Q$ is the CR spectrum injected per unit time and per unit volume (see section \ref{Injection}).
The first term on the right hand side of Eq.~(\ref{CRs_transport}) represents the spatial diffusion of CRs through the outflow depending on the turbulence and strength of the magnetic field (see section \ref{Diffusion}) and spatial advection which acts to suppress the CR flux. The second term represents momentum advection which moves the CRs to lower energies as they are transported in the expanding advective outflow. The fourth term is the CR loss rate which is the rate at which the CRs lose their energy. A hadronic CR experiences energy losses through inelastic $pp$ collisions \citep{2007Ap&SS.309..365G} in the halo region with a loss time scale $\tau_{\rm loss}\approx 20$~Gyr, on a halo distance scale of 10~kpc. 

For the results (see section \ref{CRs_propgation}), the CR number density, $n_{\rm CR}$, will be presented instead of $f$. In order to determine $n_{\rm CR}$ from $f$ which the transport equation provides, the momentum dependence of the phase space distribution is integrated out,
\begin{equation}\label{density_CR}
    n_{\rm CR}  = \int_{p_{\rm min}}^{p_{\rm max}} 4 \pi f p^2 dp = \int_{p_{\rm min}}^{p_{\rm max}}\frac{dN}{d^{3}x dp}dp.
\end{equation}

\noindent Here $p_{\rm min}$ = 10~GeV$/c$ and $p_{\rm max}$ = 30~GeV$/c$ are the minimum and maximum momentum for the injected CRs for the results presented in section \ref{CRs_propgation}. This range has been chosen to coincide with one of the selected ranges by \citet{Ackermann_2014} for the $\gamma$-ray observations of the Fermi bubbles.

\subsection{Diffusion coefficient}\label{Diffusion}

The diffusion coefficient determines the distance that a CR, on average, can travel through ambient magnetic fields until its momentum vector has changed significantly ($\sim$1~rad. in angle). For the purpose of the calculations, a simple commonly used expression for the diffusion coefficient is adopted \citep{Schlickeiser_1989, Jokipii_1996}, which expressed as a length-scale is,
\begin{equation}
\frac{D}{c} = 0.1 \left(\frac{p}{10\ \text{GeV/c}}\right)^{2-\gamma}~\text{pc},
\label{diffusion}
\end{equation}
\noindent where $c$ is the speed of light and $\gamma=5/3$ is the Kolmogorov turbulence power-spectrum slope.

\subsection{Injection and CR spectrum}\label{Injection}

The injection site for the CR source is assumed to be at the base of the outflow at $r_0$. The volumetric injection rate $Q$ is defined as,
\begin{equation}
Q=\frac{d\dot{N}}{d^{3}x dp},
\label{source}
\end{equation}
where $\dot{N}$ is the number of particles injected per unit time. 

A continuous injection spectrum has been used such that $\frac{d\dot{N}}{dp} \propto p^{-\alpha}\exp(-p/p_{\rm max})$, with $\alpha=2$ representative of diffusive shock acceleration \citep{Bell_1978}, and  $p_{\rm max}=10^{7}$~GeV$/c$. The total injected luminosity is expressed as
\begin{equation}
L_{\rm CR} = c\int^{p_{\rm max}}_{p_{\rm min}} p\frac{d\dot{N}}{dp}dp,
\label{CR_luminosity}
\end{equation}
The total injected luminosity has been set to $L_{\rm CR} = 6\times 10^{39}$~erg~s$^{-1}$ which determines the value of $\dot{N}$. This value has been chosen to match with the fitting range for the energy flux provided by the observations of the Fermi bubbles \citep{Ackermann_2014}.

\subsection{Numerical setup}\label{CR_setup}

Based on a 2D differencing scheme \citep{2017MNRAS.472...26R,2020MNRAS.499.2124R}, the temporal evolution of $f$, using a steady-state subsonic velocity solution for the advection terms, is obtained for a timescale of 300~Myr. This timescale corresponds to the time needed to reach a steady-state out to 20 kpc, the spatial region of interest in this work. The grid adopted in these calculations has a cylindrical ($R,z$) geometry with logarithmic spacing for both $R$ and $z$ grids with 350 bins. The outflow velocity, $v$, obtained from the hydrodynamic simulations (see section \ref{Galactic_model}) has been used as an input for Eq.~(\ref{CRs_transport}). Considering this velocity profile (see section \ref{Pluto_setup}) the inner boundary has been setup at 300~pc, and the outer boundary at 100~kpc. The CRs are injected close to the inner boundary. The momentum grid is logarithmically spaced with 5 bins ranging from $10-30$~GeV$/c$. The simulations have been performed for one quadrant only. The results presented here have subsequently been reflected into the other quadrants.

\section{Results}
\label{Results}

\subsection{Hydrodynamic outflow}\label{Hydrodynamic outflow}

\begin{figure}
\includegraphics[width=0.5\textwidth]{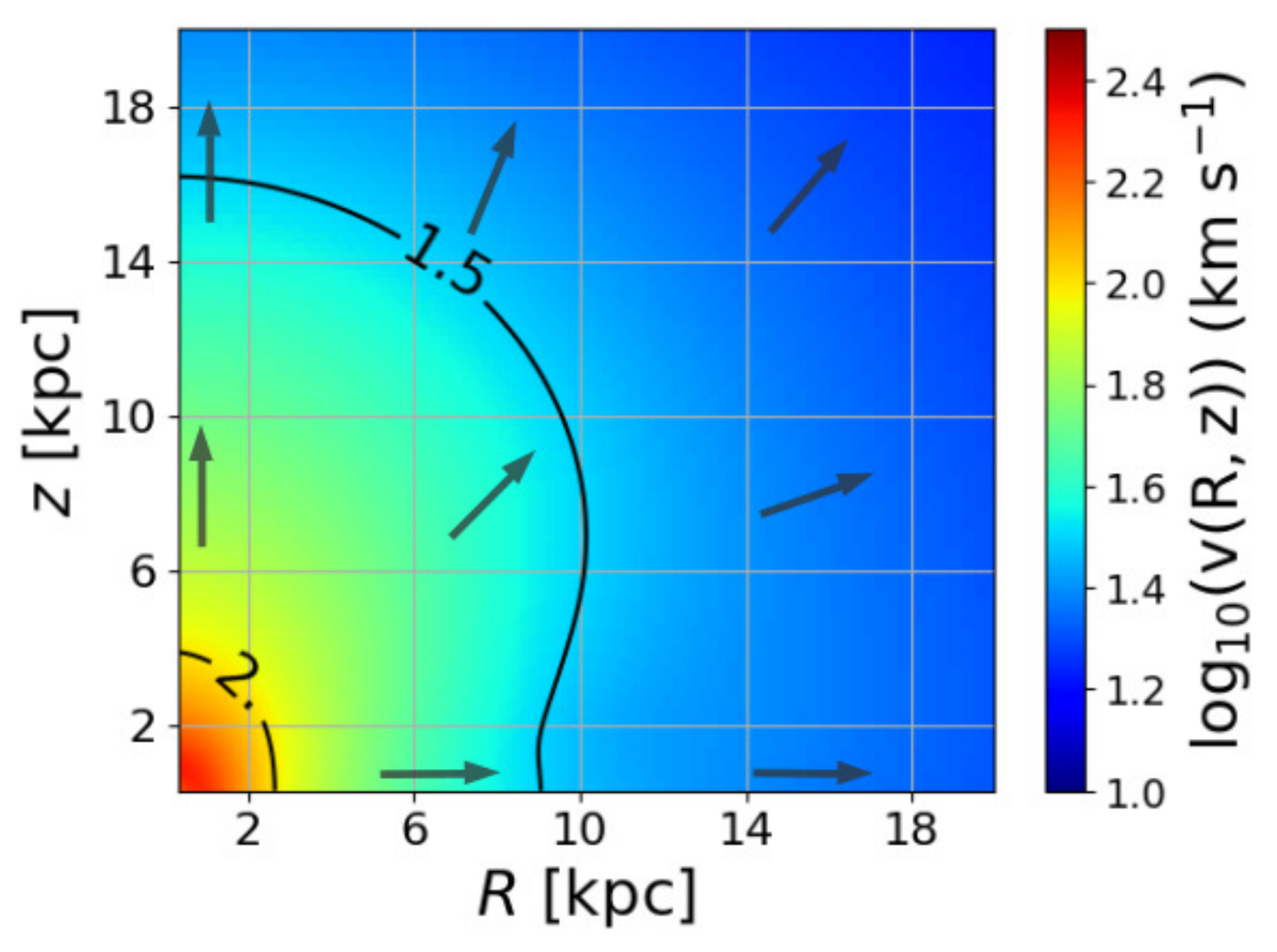}
\caption{2D spatial distribution of the subsonic velocity profile through a gravitational potential composed of a bulge, disc and halo with $c_s \approx$ 250 km s$^{-1}$. The arrows represent the direction of propagation of the outflow. The continuous colourbar indicates the logarithm (base 10) of the magnitude of the velocity.}
\label{Rotation}
\end{figure}

Fig.~\ref{Rotation} shows the 2D velocity profile of the subsonic outflow in both the $R$ and $z$-directions. The Galactic potential adopted in this paper (see section \ref{potential}), in combination with a temperature of $kT\approx 400~\rm{eV}$ for the halo region, give two roots for Eq.~(\ref{outflow}). The velocity profile along the $z$-axis is shown in Fig. \ref{Velocity_comparison}. Once injected at the inner radius (0.3~kpc) the gas first decelerates until reaching a distance of $\sim0.45~\rm{kpc}$ (see Appendix \ref{Appendix_B}) and then accelerates. This acceleration continues outwards up to $r_c$$\,\approx\,$$1~\rm{kpc}$ (red region in Fig.~\ref{Rotation}), reaching a velocity of $v \approx 210~{\rm km~s}^{-1}$, after which the gas then decelerates continuously (as shown by the orange to blue regions in Fig.~\ref{Rotation}). The outflow spreads mainly along the z-direction due to the asphericity of the disc component of the Galactic potential, see section \ref{1D_outflow}. Near the Galactic center, the outflow takes a conic shape and extends progressively to a lobular shape as the outflow decelerates. It is found that the centrifugal force due to Galactic rotation has only a minor effect on the results. It reduces the gravitational force acting on the outflow at larger distances from the rotation axis slightly and mildly enhances the extension of the outflow velocity in the $R$ direction. 
The continuous deceleration beyond $r_c$, along increasing Galactic latitude, gives the following velocities: $v\,(z=3~{\rm kpc}) \approx  120~{\rm km s}^{-1}$, $v\,(z=5~{\rm kpc}) \approx 85~{\rm km~s}^{-1}$, $v\,(z=7~{\rm kpc}) \approx 70~{\rm km~s}^{-1}$ and $v\,(z=10~{\rm kpc}) \approx 53~{\rm km~s}^{-1}$. These results appear compatible with observations of UV absorption lines of OI, AlII, CII, CIV, SiII, SiIII, SiIV and other species provided by \citet{2017ApJ...834..191B} for the Local Standard of Rest velocity. However, the velocities given above are lower than the values inferred from similar UV absorption line studies \citep{2018ApJ...860...98K} for the South Fermi bubbles. It should be noted that an asymmetry between the North and South Fermi lobes exists, potentially related to different density distributions in the two hemispheres \citep{2019MNRAS.482.4813S}. More recent observational results for both North and South Fermi lobes provided by \citet{2020ApJ...898..128A} give a outflow profile similar to \citet{2018ApJ...860...98K} but with a somewhat faster velocity, by a factor $\sim10$.

\subsection{CR density map}\label{CRs_propgation}
%\subsubsection{CRs spatial density distribution}\label{CRs spatial density distribution}

\begin{figure}
\includegraphics[width=0.5\textwidth]{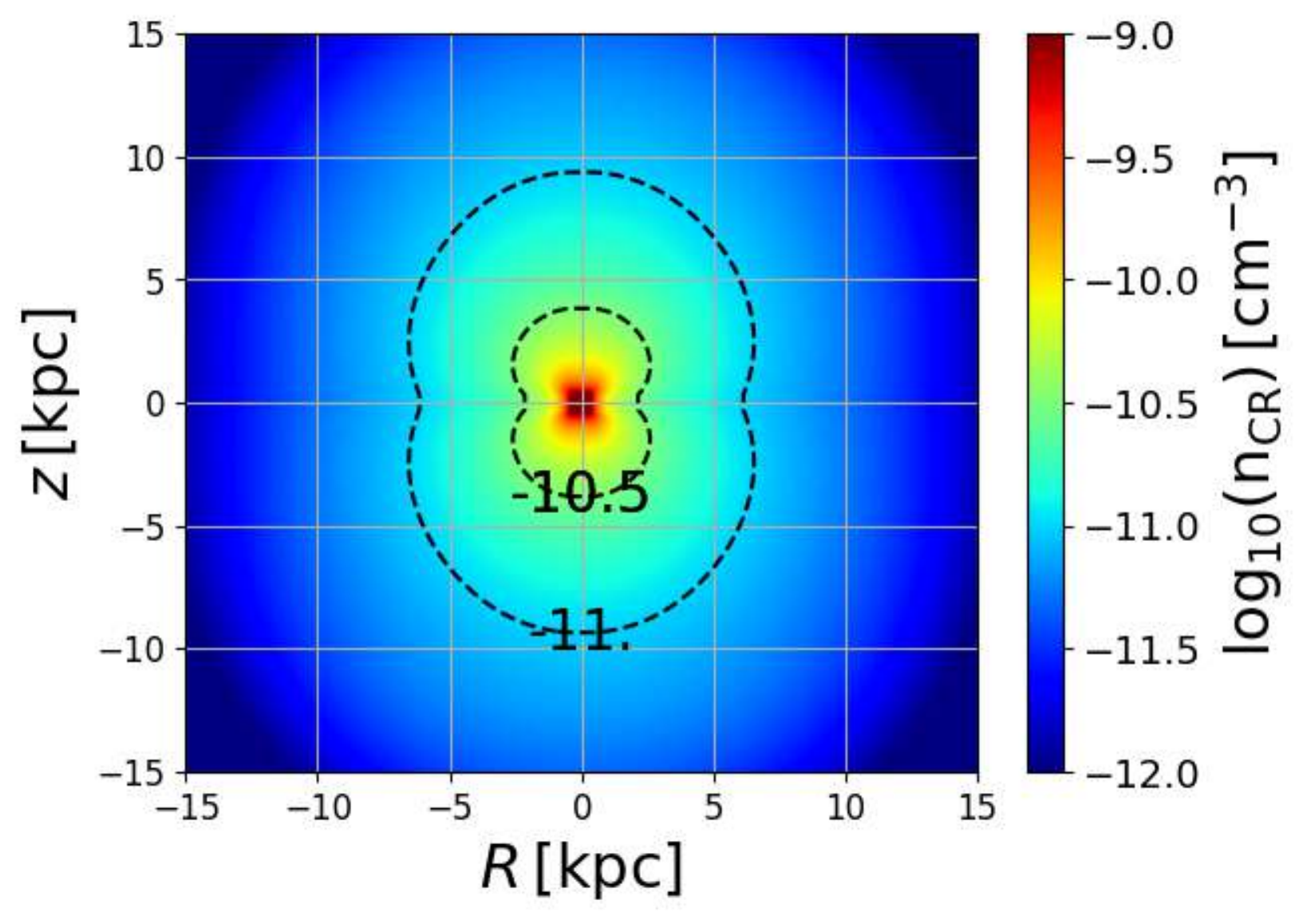}
\caption{2D CR spatial density distribution in the Galaxy for a energy range of E$_{\rm{CR}}$ = 10-30~GeV. The dashed black lines are contours representing different density distribution. The continuous colourbar indicates the logarithm (base 10) of the cosmic ray density.}
\label{Number_density}
\end{figure}

Here the results related to the CR propagation through the outflow velocity region described in section \ref{Hydrodynamic outflow} are presented. Advection and diffusion are the two terms in the transport equation (Eq.~(\ref{CRs_transport})) which compete to dictate the resultant CR transport. The diffusion mechanism can be ignored if the outflow velocity is high enough, and one is considering low energy ($< 10$~GeV) CRs (see Eq.~\ref{diffusion}). However for the case of a Galactic breeze, $v$ is sufficiently small such that both terms play an important role in dictating the resultant CR spatial distribution. This effect can be seen in Fig.~\ref{Number_density}, which shows a 2D plot of the CR number density distribution in the Galaxy produced with the model. In the vicinity of the GC, where the outflow velocity is the highest (as shown in by the red to blue-green region in Fig.~\ref{Number_density}), advection is the dominant transport mechanism. In this region, due to the asymmetry of the Galactic potential, the CRs propagate mainly along the $z$-direction. The velocity difference in the $R$ and $z$-direction is less pronounced as the outflow decelerates. Subsequently, the bubble-shape becomes more spherical in shape at higher latitudes.

\subsection{Gamma-ray emission map}\label{gamma_ray_emission_map}

Inelastic CR collisions with ambient gas in the Galactic halo results in the creation of charged and neutral pions. The dominant decay mode for neutral pions is into two photons \citep{2020PTEP.2020h3C01P}. A photon produced by such a collision obtains $\sim 10~\%$ of the initial CR proton energy. Therefore, for the 10-30~GeV CR energy range considered, photons with energy $E_{\gamma}$ = 1 - 3 GeV are produced. The total resulting gamma-ray brightness, $E_{\gamma}F_{\gamma}$, seen by an observer along a particular solid angle of angular size $\Delta \Omega$ is obtained from summing the contribution over the full line-of-sight,

\begin{equation}\label{gamma_flux}
 E_{\gamma}F_{\gamma}=\frac{1}{3\Delta \Omega} \int \frac{dL_{\gamma}}{4\pi r_{\rm obs}^{2}} = \frac{1}{12\pi \tau_{\rm loss}} \int e_{CR}dr_{\rm obs},
\end{equation}
\noindent where $r_{\rm obs}$ is the distance between the source of emission and the observer. The factor of $\frac{1}{3}$ in the above expression accounts for the fraction of neutral pions produced in inelastic CR collisions with the target gas. These neutral pions, $\pi^0$, subsequently quickly decay to gamma-ray photons, whereas the other two charged pions, $\pi^{+}$ and $\pi^{-}$, also produced do not decay directly into gamma-ray photons. The position of the observer from the GC has been fixed at 8~kpc \citep{2017ApJ...837...30G}. The CR energy density, $e_{\rm CR}$ is expressed as
\begin{equation}
    e_{\rm{CR}} = \int_{p_{\rm min}}^{p_{\rm max}} 4 \pi f p^3 dp.
\end{equation}

\begin{figure*}
\includegraphics[width=0.5\linewidth]{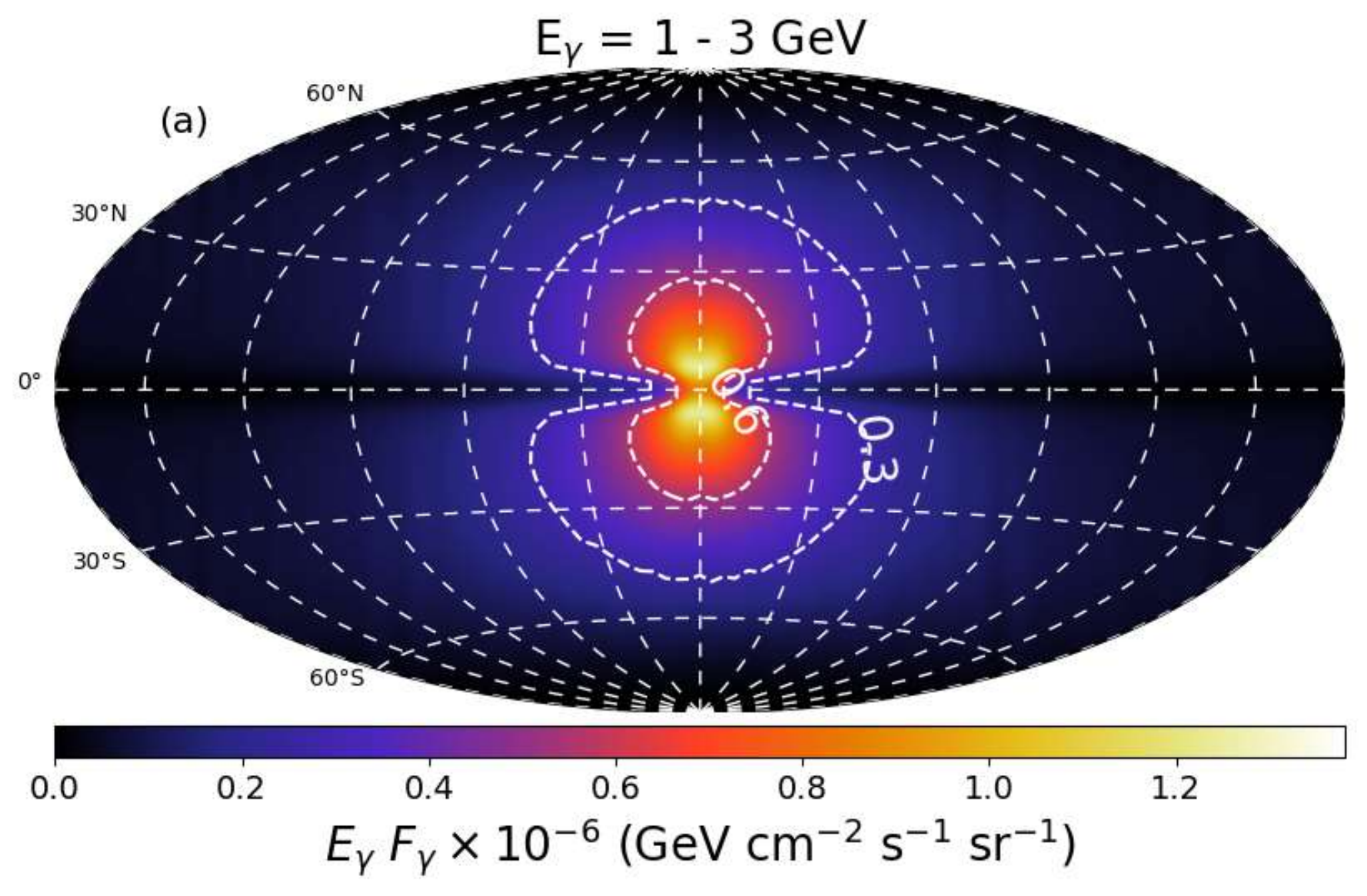}%
\includegraphics[width=0.5\linewidth]{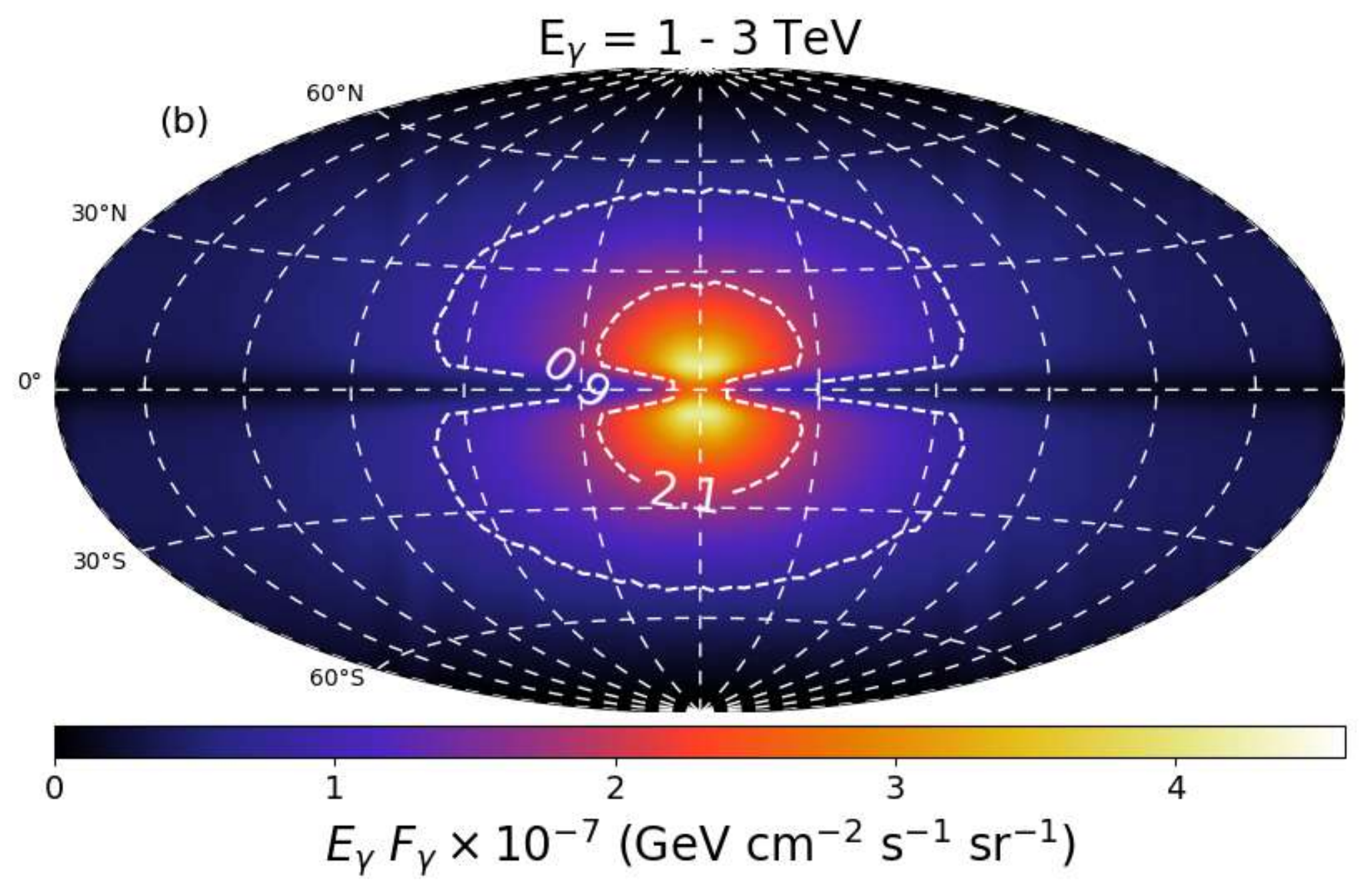}
\caption{$\gamma$-ray emission maps produced with a subsonic outflow. \textbf{Left:} The emission is produced by photons with energy ranging from 1 to 3 GeV. \textbf{Right:} The emission is produced by photons with energy ranging from 1 to 3 TeV. It is important to note that colour bar limits are not the same for the two plots.}
\label{gamma_map}
\end{figure*}

The energy flux from all of the cells contained along the line-of-sight within each solid angle direction is summed up. The skymap of the gamma-ray energy flux density is shown in Fig.~\ref{gamma_map}a for $E_{\gamma} = 1-3~\rm{GeV}$ (Fig.~\ref{gamma_map}b is for $E_{\gamma} = 1-3~\rm{TeV}$, see section \ref{CTA}). The simulated subsonic velocity profile in combination with $L_{\rm CR} = 6\times 10^{39}~{\rm erg~s}^{-1}$ produces bilobal gamma-ray emission, with a pronounced pinch at low Galactic latitudes. From observations, the height of the Fermi bubbles is $\sim\,$10~kpc, corresponding to $\sim\,$50$^{\circ}$ for an observer at $\sim\,$8~kpc. For the gamma-ray energy range, $E_{\gamma} = 1-3~\rm{GeV}$ produced with a subsonic outflow, the total gamma-ray luminosity is $L_{\gamma} \approx 7 \times 10^{36} \rm{erg~s^{-1}}$. In comparison, observations indicate a gamma-ray luminosity, for an energy range from 0.1-500 GeV to be $L_{\gamma} \approx 4.4\times 10^{37}~\rm{erg~s^{-1}}$ \citep{Ackermann_2014}, assuming a fixed emitting distance of 8~kpc. Adjusting their luminosity to our reduced energy range of 1-3~GeV, leads to $L_{\gamma} \approx 7.5\times 10^{36}~\rm{erg~s^{-1}}$, which agrees well with the gamma-ray luminosity from our simulation.

The brightness of the gamma-ray energy flux produced appears broadly consistent with Fermi-LAT observations. However, the width of the observed Fermi bubbles, estimated to be around $20^{\circ}$ from the central axis  \citep{Ackermann_2014} are narrower than the simulation results shown here, which extend to a width of $30^{\circ}$. This difference originates from the assumptions in the adopted setup. Specifically, both the magnitude of the CR diffusion coefficient (see Eq.~\ref{CRs_transport}), and the isothermal gas temperature, collectively dictate the subsequent lobe width.

\subsubsection{Predictions for CTA and SWGO}\label{CTA}

The Cherenkov Telescope array \citep[CTA,][]{CTA} and the Southern Wide-field gamma-ray Observatory \citep[SWGO,][]{SWGO}, a wide field observatory that will complement CTA, are next generation ground-base gamma-ray instruments, which collectively will span an energy range from 20~GeV to 300~TeV \citep{2021JCAP...02..048A}. A simulation for the transport of $E=10-30$~TeV energy CRs has been run in order to predict what future observations of the Fermi bubbles will look like if a subsonic outflow is responsible for the gamma-ray emission. The gamma-ray emission map produced by the CRs is shown in Fig.~\ref{gamma_map}b. These simulations are done with the same setup, for both the hydrodynamic outflow simulation (see Section\,\ref{Pluto_setup}) and the CR transport code (see Section\,\ref{CR_setup}). 

\begin{figure*}
\includegraphics[width=0.5\linewidth]{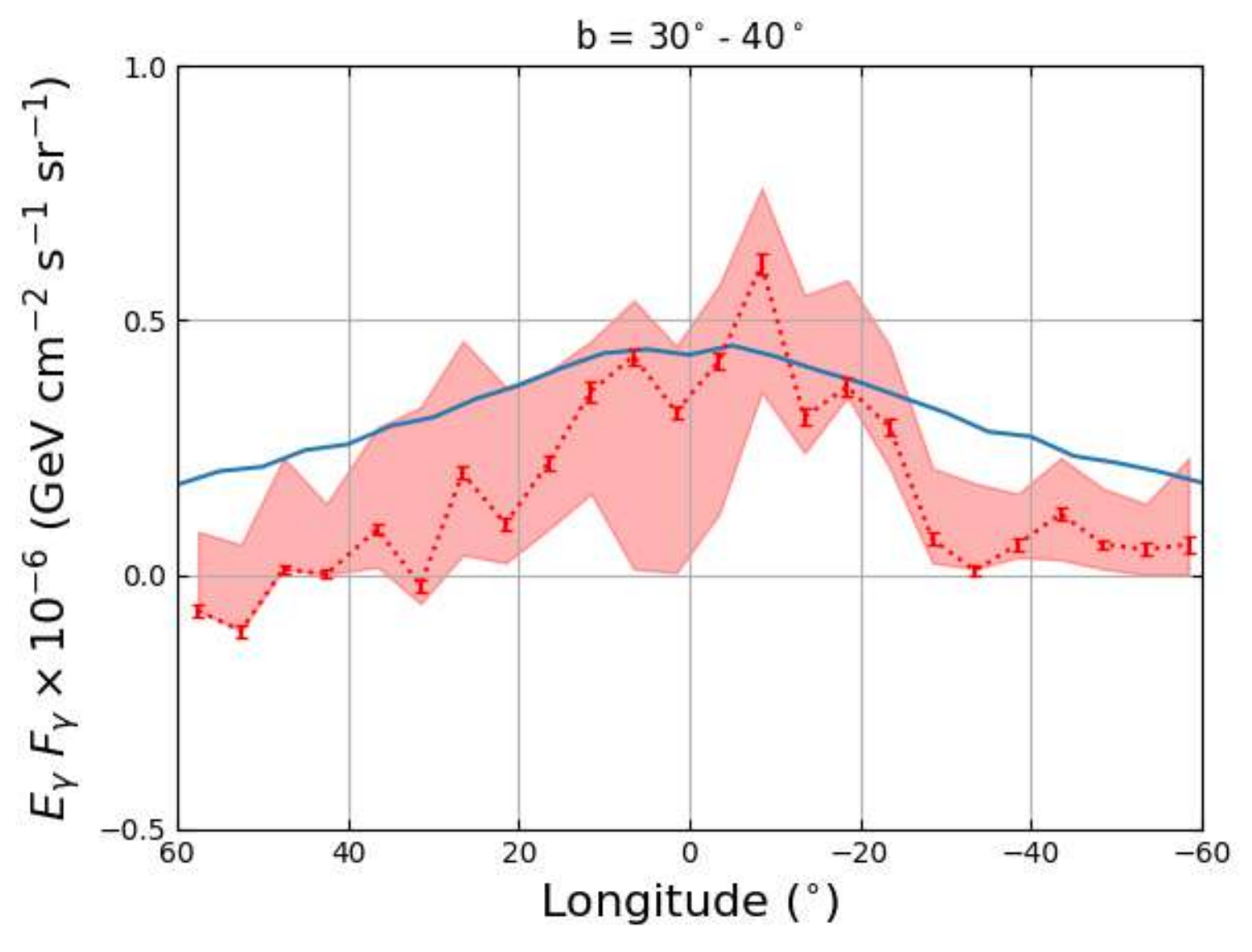}%
\includegraphics[width=0.5\linewidth]{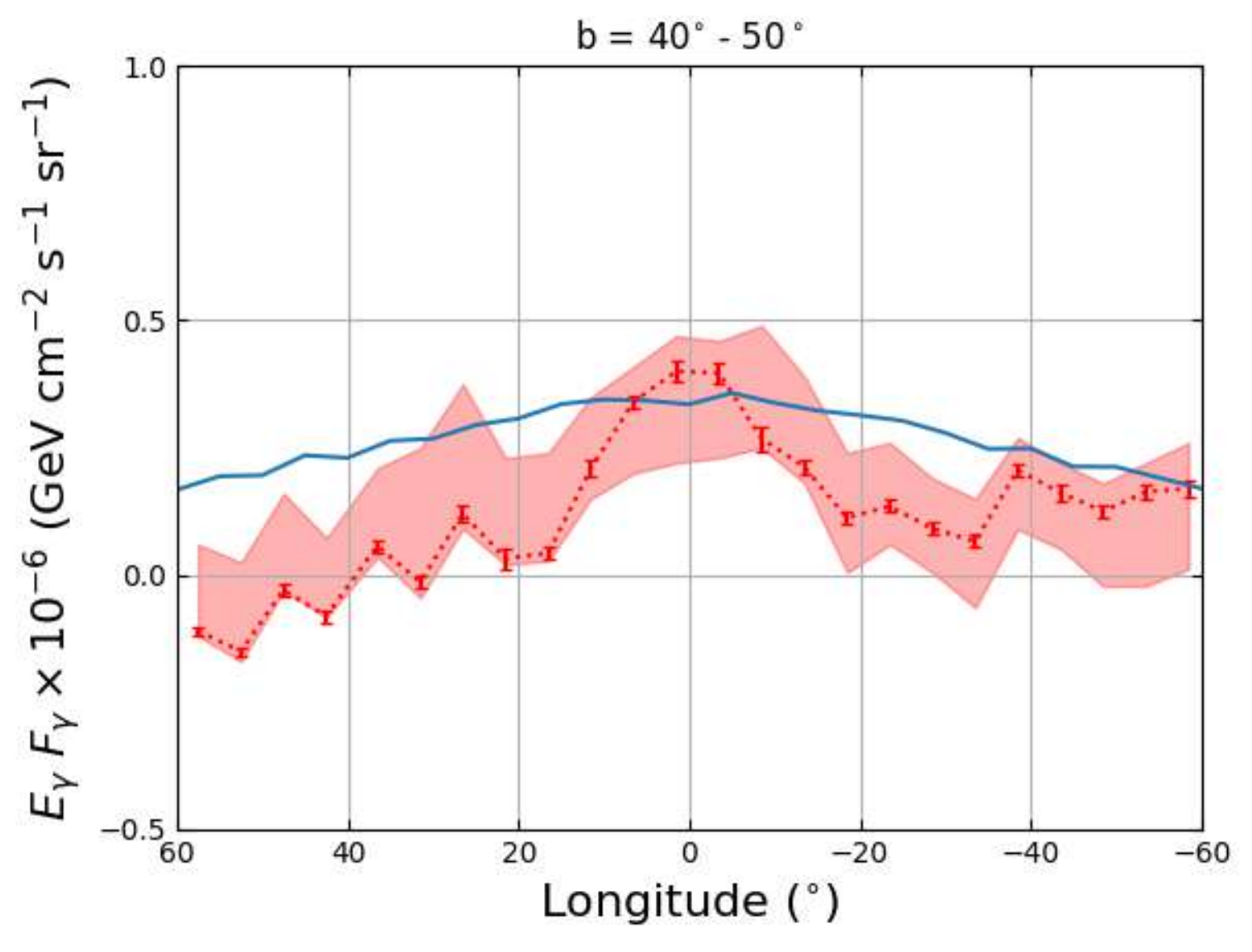}
\caption{The photon energy flux produced by the gamma-ray emission is plotted as a function of Galactic longitude for different latitude ranges. The blue line represent the energy flux provided by the simulations for a subsonic outflow presented in this paper. The error bar ranges and the shaded red region are provided by the Fermi-LAT observational results \citep{Ackermann_2014}. The red error bars represent the energy flux determined from observations of the Fermi bubbles. The shaded red region has been computed from uncertainties of different models and different possible templates. \textbf{Left:} The photon energy flux is compared with Fermi-LAT data for $b = 30^{\circ}-40^{\circ}$. \textbf{Right:} The photon energy flux is compared with Fermi-LAT data for $b = 40^{\circ}-50^{\circ}$. }
\label{Comparison}
\end{figure*}

Considering the adopted subsonic velocity profile and the high CR energies, diffusion plays a more prominent role (see Eq.~(\ref{diffusion})). Compared to the lower energy gamma-ray emission map shown in Fig.~\ref{gamma_map}a, the gamma-ray distribution is wider and reaches a slightly lower height. This difference is due to the energy dependence of the diffusion coefficient adopted, which introduces energy dependent transport effects. For a latitude of $\sim 50^{\circ}$ and a longitude of $\sim 0^{\circ}$ the predicted energy flux observed, according to the subsonic model presented here, should be $\sim 0.9\times 10^{-8}\,\rm{GeV}\,\rm{cm}^{-2}\,\rm{s}^{-1}\,\rm{sr}^{-1}$. \\
%
%FORCE a new line here
\newline
Returning to the comparison between the Fermi-LAT data and Galactic breeze simulations results, in Fig.~\ref{Comparison} a direct comparison is made between the simulations and data provided by the Fermi-LAT collaboration \citep{Ackermann_2014} for latitudes from $b = 30^{\circ} - 50^{\circ}$. The blue line represents the energy flux provided by the simulations for a subsonic outflow. The red error bar ranges come from the data defined with the GALPROP CR propagation and interactions code \citep{Galprop} as templates. The red shaded regions were computed from uncertainties of different models and different definitions of the templates. Both the red error bar ranges and the red shaded regions are from \citet{Ackermann_2014}.  At latitudes between $30^{\circ}<b< 40^{\circ}$ and $40^{\circ}< b < 50^{\circ}$, (for -15$^{\circ}<l< 15^{\circ}$), the energy flux is broadly compatible with the observations, although as already noted previously, beyond a longitude of 15$^{\circ}$ the energy flux no longer falls within the red shaded region due to the broader width of the simulated lobes. 

\section{Conclusions and Outlook}
\label{Conclusions}
%\subsection{Gravitational potential and density distribution}

 Following on from an earlier investigation \citep{2017PhRvD..95b3001T} into a subsonic outflow origin for the Fermi bubbles, a hydrodynamical outflow simulation has been carried out here. The outflow velocity profile provided by the hydrodynamic simulation is then used as part of a 2D CR transport code description. CRs transported by the outflow into the Galactic halo region subsequently undergo inelastic $pp$ energy loss interactions with the low density hot ambient gas present. These energy loss interactions give rise to spatially dependent gamma-ray emission. A comparison of this emission is made, for a specific photon energy range of $1-3~\rm{GeV}$, with Fermi bubbles observations data provided by the Fermi-LAT satellite. Following these results a higher energy emission map prediction is provided, in the $1-3~\rm{TeV}$ energy range of relevance for next generation instruments (CTA/SWGO). 

 The hydrodynamic simulation results presented in this paper reveal that the Galactic breeze propagates predominantly in the direction orthogonal to the Galactic plane. This outflow subsequently develops a conical shape, which eventually become lobe-like in shape at large distances. The continuous deceleration of the outflow beyond the critical radius appears consistent with UV absorption line observations of cold clouds within the Fermi bubbles region  \citep{2017ApJ...834..191B,2020ApJ...898..128A}. 
Moreover, the Galactic breeze model does not sweep up the ambient Galactic density as winds do. The initial hydrostatic density distribution is instead largely preserved, motivating halo densities which appear consistent with observations provided for the Milky-Way \citep{2017ApJ...836..243G,Zhang_2021,Martynenko_2022}.
A comparison of the gamma-ray skymap emission produced by the CR energy losses show that these are broadly compatible with the Fermi bubbles observations provided by \citet{Ackermann_2014}. This compatibility obtained from a model with so few assumptions supports the proposal that a Galactic breeze provides a self-consistent description for the origin of the bubbles. However it is noted that further improvements to the model are motivated. Specifically, these improvements concern the adoption of a more complex temperature distribution and the inclusion of magnetic fields in the PLUTO simulation.

Fundamentally, the present limited knowledge about the nature of the Galactic halo region (ie. the gas density and temperature, and the shape of the Galactic potential) limits the ability to describe gas and CR transport through this region. Of particular additional concern are Galactic halo magnetic fields, which have not been considered explicitly in the transport description used here. However, these fields certainly play an intrinsic role in CR diffusion, and likely also play a role in a more complete magneto-hydrodynamic description of the gas outflow. Furthermore, the presence and strength of these magnetic fields may well underlie aspects of the Fermi bubbles morphology. For example, the bubbles edges observed in gamma-ray emission may relate to the magnetic field geometry in this region, hindering the diffusion of particles across the bubbles edges. This would provide a dimmer surface brightness at the bubbles edges.

Additional information about the Galactic halo magnetic fields have been provided by the WMAP Haze \citep{2008ApJ...680.1222D,2012ApJ...750...17D} and S-PASS bubble \citep{2013Natur.493...66C}. Both the  WMAP Haze and S-PASS bubble appear spatially coincident with the base of the Fermi bubbles, although the haze appears to extend only up to $\sim$~6~kpc in height. Within the Galactic breeze scenario, this synchrotron emission is produced by both primary electrons, and secondary electrons produced through pp interactions, with the former dominating at low Galactic latitudes. Future PLUTO simulations incorporating magnetic fields will allow this possibility to be explored.

The eROSITA soft X-ray bubbles also provides new insights about the Galactic halo environment. This emission is thought to originate from energetic thermal electrons, indicating that a heating mechanism for this gas is operating. Within the Galactic breeze thermally driven outflow model, this same heating mechanism is responsible for the hot isothermal gas temperature adopted. Potentially, the heating mechanism for the gas may be magnetic (Alfvén wave) heating. However, provided that a hot gas temperature is maintained, the driving mechanism for the heating of the gas is not of primary focus.

Further observations probing the nature of the halo environment are needed in order to provide a better constraint on the Galactic potential and temperature for the gas throughout the Galaxy. Observations of cold clouds in the Fermi bubbles, at 2$\,-\,$3~kpc from the GC, give a velocity outflow of $\sim\,$330~km~s$^{-1}$ \citep{2020ApJ...898..128A}. Such a value is hardly reachable for an isothermal subsonic description considering the fitting range for the Galactic gravitational potential provided by \citet{2019ApJ...873..118W}. On the other hand, the apparent continuous deceleration of the gas observed with increasing latitude \citep{2017ApJ...834..191B, 2020ApJ...898..128A}, is compatible with a subsonic outflow scenario. A single temperature model, therefore, may provide an insufficiently accurate account of the gas velocity evolution. Indeed, evidence for a transition between a hot bulge and an even hotter halo ($\sim 1$~keV) out at larger distances \citep{2019ApJ...882L..23D, 2019ApJ...887..257D, Das_2021}, motivates the consideration of a more complex temperature profile than the one considered here.

\section*{Acknowledgements}
The authors would like to thank the anonymous reviewer for their valuable comments which helped to
improve the manuscript.
OT and AMT acknowledge support from DESY (Zeuthen, Germany), a member of the Helmholtz Association HGF. DRL would like to acknowledge that this publication has emanated from research conducted with the financial support of Science Foundation Ireland under Grant number 21/PATH-S/9339.

\section*{Data Availability}
The data underlying this article will be shared on reasonable request to the corresponding author.

\bibliographystyle{mnras}
\bibliography{sample.bib}

\begin{thebibliography}{}
\makeatletter
\relax
\def\mn@urlcharsother{\let\do\@makeother \do\$\do\&\do\#\do\^\do\_\do\%\do\~}
\def\mn@doi{\begingroup\mn@urlcharsother \@ifnextchar [ {\mn@doi@}
  {\mn@doi@[]}}
\def\mn@doi@[#1]#2{\def\@tempa{#1}\ifx\@tempa\@empty \href
  {http://dx.doi.org/#2} {doi:#2}\else \href {http://dx.doi.org/#2} {#1}\fi
  \endgroup}
\def\mn@eprint#1#2{\mn@eprint@#1:#2::\@nil}
\def\mn@eprint@arXiv#1{\href {http://arxiv.org/abs/#1} {{\tt arXiv:#1}}}
\def\mn@eprint@dblp#1{\href {http://dblp.uni-trier.de/rec/bibtex/#1.xml}
  {dblp:#1}}
\def\mn@eprint@#1:#2:#3:#4\@nil{\def\@tempa {#1}\def\@tempb {#2}\def\@tempc
  {#3}\ifx \@tempc \@empty \let \@tempc \@tempb \let \@tempb \@tempa \fi \ifx
  \@tempb \@empty \def\@tempb {arXiv}\fi \@ifundefined
  {mn@eprint@\@tempb}{\@tempb:\@tempc}{\expandafter \expandafter \csname
  mn@eprint@\@tempb\endcsname \expandafter{\@tempc}}}

\bibitem[\protect\citeauthoryear{{Abdalla} et~al.,}{{Abdalla}
  et~al.}{2021}]{2021JCAP...02..048A}
{Abdalla} H.,  et~al., 2021, \mn@doi [\jcap] {10.1088/1475-7516/2021/02/048},
  \href {https://ui.adsabs.harvard.edu/abs/2021JCAP...02..048A} {2021, 048}

\bibitem[\protect\citeauthoryear{Abreu et~al.}{Abreu et~al.}{2019}]{SWGO}
Abreu P.,  et~al., 2019, {The Southern Wide-Field Gamma-Ray Observatory (SWGO):
  A Next-Generation Ground-Based Survey Instrument for VHE Gamma-Ray
  Astronomy}.
 (\mn@eprint {arXiv} {1907.07737})

\bibitem[\protect\citeauthoryear{Acharya et~al.}{Acharya et~al.}{2018}]{CTA}
Acharya B.~S.,  et~al., 2018, {Science with the Cherenkov Telescope Array}.
WSP (\mn@eprint {arXiv} {1709.07997}), \mn@doi{10.1142/10986}

\bibitem[\protect\citeauthoryear{{Ackermann} et~al.,}{{Ackermann}
  et~al.}{2014}]{Ackermann_2014}
{Ackermann} M.,  et~al., 2014, \mn@doi [\apj] {10.1088/0004-637X/793/1/64},
  \href {https://ui.adsabs.harvard.edu/abs/2014ApJ...793...64A} {793, 64}

\bibitem[\protect\citeauthoryear{{Ashley}, {Fox}, {Jenkins}, {Wakker},
  {Bordoloi}, {Lockman}, {Savage}  \& {Karim}}{{Ashley}
  et~al.}{2020}]{2020ApJ...898..128A}
{Ashley} T.,  {Fox} A.~J.,  {Jenkins} E.~B.,  {Wakker} B.~P.,  {Bordoloi} R.,
  {Lockman} F.~J.,  {Savage} B.~D.,   {Karim} T.,  2020, \mn@doi [\apj]
  {10.3847/1538-4357/ab9ff8}, \href
  {https://ui.adsabs.harvard.edu/abs/2020ApJ...898..128A} {898, 128}

\bibitem[\protect\citeauthoryear{{Bell}}{{Bell}}{1978}]{Bell_1978}
{Bell} A.~R.,  1978, \mn@doi [\mnras] {10.1093/mnras/182.2.147}, \href
  {https://ui.adsabs.harvard.edu/abs/1978MNRAS.182..147B} {182, 147}

\bibitem[\protect\citeauthoryear{{Bhattacharjee}, {Chaudhury}  \&
  {Kundu}}{{Bhattacharjee} et~al.}{2014}]{2014ApJ...785...63B}
{Bhattacharjee} P.,  {Chaudhury} S.,   {Kundu} S.,  2014, \mn@doi [\apj]
  {10.1088/0004-637X/785/1/63}, \href
  {https://ui.adsabs.harvard.edu/abs/2014ApJ...785...63B} {785, 63}

\bibitem[\protect\citeauthoryear{{Bordoloi} et~al.,}{{Bordoloi}
  et~al.}{2017}]{2017ApJ...834..191B}
{Bordoloi} R.,  et~al., 2017, \mn@doi [\apj] {10.3847/1538-4357/834/2/191},
  \href {https://ui.adsabs.harvard.edu/abs/2017ApJ...834..191B} {834, 191}

\bibitem[\protect\citeauthoryear{{Burbidge}, {Burbidge}  \& {Rubin}}{{Burbidge}
  et~al.}{1964}]{1964ApJ...140..942B}
{Burbidge} E.~M.,  {Burbidge} G.~R.,   {Rubin} V.~C.,  1964, \mn@doi [\apj]
  {10.1086/147997}, \href
  {https://ui.adsabs.harvard.edu/abs/1964ApJ...140..942B} {140, 942}

\bibitem[\protect\citeauthoryear{{Bustard}, {Pardy}, {D'Onghia}, {Zweibel}  \&
  {Gallagher}}{{Bustard} et~al.}{2018}]{Bustard2018}
{Bustard} C.,  {Pardy} S.~A.,  {D'Onghia} E.,  {Zweibel} E.~G.,   {Gallagher}
  J.~S. I.,  2018, \mn@doi [\apj] {10.3847/1538-4357/aad08f}, \href
  {https://ui.adsabs.harvard.edu/abs/2018ApJ...863...49B} {863, 49}

\bibitem[\protect\citeauthoryear{{Carretti} et~al.,}{{Carretti}
  et~al.}{2013}]{2013Natur.493...66C}
{Carretti} E.,  et~al., 2013, \mn@doi [\nat] {10.1038/nature11734}, \href
  {https://ui.adsabs.harvard.edu/abs/2013Natur.493...66C} {493, 66}

\bibitem[\protect\citeauthoryear{{Cashman} et~al.,}{{Cashman}
  et~al.}{2021}]{2021ApJ...923L..11C}
{Cashman} F.~H.,  et~al., 2021, \mn@doi [\apjl] {10.3847/2041-8213/ac3cbc},
  \href {https://ui.adsabs.harvard.edu/abs/2021ApJ...923L..11C} {923, L11}

\bibitem[\protect\citeauthoryear{{Chamberlain}}{{Chamberlain}}{1965}]{1965ApJ...141..320C}
{Chamberlain} J.~W.,  1965, \mn@doi [\apj] {10.1086/148119}, \href
  {https://ui.adsabs.harvard.edu/abs/1965ApJ...141..320C} {141, 320}

\bibitem[\protect\citeauthoryear{{Chan}, {Keres}, {Gurvich}, {Hopkins},
  {Trapp}, {Ji}  \& {Faucher-Giguere}}{{Chan} et~al.}{2021}]{Chan2021}
{Chan} T.~K.,  {Keres} D.,  {Gurvich} A.~B.,  {Hopkins} P.,  {Trapp} C.,  {Ji}
  S.,   {Faucher-Giguere} C.-A.,  2021, arXiv e-prints, \href
  {https://ui.adsabs.harvard.edu/abs/2021arXiv211006231C} {p. arXiv:2110.06231}

\bibitem[\protect\citeauthoryear{{Cheng} et~al.,}{{Cheng}
  et~al.}{1997}]{1997ApJ...481L..43C}
{Cheng} L.~X.,  et~al., 1997, \mn@doi [\apjl] {10.1086/310638}, \href
  {https://ui.adsabs.harvard.edu/abs/1997ApJ...481L..43C} {481, L43}

\bibitem[\protect\citeauthoryear{{Cheng}, {Chernyshov}, {Dogiel}, {Ko}  \&
  {Ip}}{{Cheng} et~al.}{2011}]{2011ApJ...731L..17C}
{Cheng} K.~S.,  {Chernyshov} D.~O.,  {Dogiel} V.~A.,  {Ko} C.~M.,   {Ip} W.~H.,
   2011, \mn@doi [\apjl] {10.1088/2041-8205/731/1/L17}, \href
  {https://ui.adsabs.harvard.edu/abs/2011ApJ...731L..17C} {731, L17}

\bibitem[\protect\citeauthoryear{{Chevalier} \& {Clegg}}{{Chevalier} \&
  {Clegg}}{1985}]{1985Natur.317...44C}
{Chevalier} R.~A.,  {Clegg} A.~W.,  1985, \mn@doi [\nat] {10.1038/317044a0},
  \href {https://ui.adsabs.harvard.edu/abs/1985Natur.317...44C} {317, 44}

\bibitem[\protect\citeauthoryear{{Crocker} \& {Aharonian}}{{Crocker} \&
  {Aharonian}}{2011}]{2011PhRvL.106j1102C}
{Crocker} R.~M.,  {Aharonian} F.,  2011, \mn@doi [\prl]
  {10.1103/PhysRevLett.106.101102}, \href
  {https://ui.adsabs.harvard.edu/abs/2011PhRvL.106j1102C} {106, 101102}

\bibitem[\protect\citeauthoryear{{Crocker}, {Bicknell}, {Taylor}  \&
  {Carretti}}{{Crocker} et~al.}{2015}]{2015ApJ...808..107C}
{Crocker} R.~M.,  {Bicknell} G.~V.,  {Taylor} A.~M.,   {Carretti} E.,  2015,
  \mn@doi [\apj] {10.1088/0004-637X/808/2/107}, \href
  {https://ui.adsabs.harvard.edu/abs/2015ApJ...808..107C} {808, 107}

\bibitem[\protect\citeauthoryear{{Das}, {Mathur}, {Nicastro}  \&
  {Krongold}}{{Das} et~al.}{2019a}]{2019ApJ...882L..23D}
{Das} S.,  {Mathur} S.,  {Nicastro} F.,   {Krongold} Y.,  2019a, \mn@doi
  [\apjl] {10.3847/2041-8213/ab3b09}, \href
  {https://ui.adsabs.harvard.edu/abs/2019ApJ...882L..23D} {882, L23}

\bibitem[\protect\citeauthoryear{{Das}, {Mathur}, {Gupta}, {Nicastro}  \&
  {Krongold}}{{Das} et~al.}{2019b}]{2019ApJ...887..257D}
{Das} S.,  {Mathur} S.,  {Gupta} A.,  {Nicastro} F.,   {Krongold} Y.,  2019b,
  \mn@doi [\apj] {10.3847/1538-4357/ab5846}, \href
  {https://ui.adsabs.harvard.edu/abs/2019ApJ...887..257D} {887, 257}

\bibitem[\protect\citeauthoryear{{Das}, {Mathur}, {Gupta}  \& {Krongold}}{{Das}
  et~al.}{2021}]{Das_2021}
{Das} S.,  {Mathur} S.,  {Gupta} A.,   {Krongold} Y.,  2021, \mn@doi [\apj]
  {10.3847/1538-4357/ac0e8e}, \href
  {https://ui.adsabs.harvard.edu/abs/2021ApJ...918...83D} {918, 83}

\bibitem[\protect\citeauthoryear{{Dobler}}{{Dobler}}{2012}]{2012ApJ...750...17D}
{Dobler} G.,  2012, \mn@doi [\apj] {10.1088/0004-637X/750/1/17}, \href
  {https://ui.adsabs.harvard.edu/abs/2012ApJ...750...17D} {750, 17}

\bibitem[\protect\citeauthoryear{{Dobler} \& {Finkbeiner}}{{Dobler} \&
  {Finkbeiner}}{2008}]{2008ApJ...680.1222D}
{Dobler} G.,  {Finkbeiner} D.~P.,  2008, \mn@doi [\apj] {10.1086/587862}, \href
  {https://ui.adsabs.harvard.edu/abs/2008ApJ...680.1222D} {680, 1222}

\bibitem[\protect\citeauthoryear{{Dobler}, {Finkbeiner}, {Cholis}, {Slatyer}
  \& {Weiner}}{{Dobler} et~al.}{2010}]{Dobler_2010}
{Dobler} G.,  {Finkbeiner} D.~P.,  {Cholis} I.,  {Slatyer} T.,   {Weiner} N.,
  2010, \mn@doi [\apj] {10.1088/0004-637X/717/2/825}, \href
  {https://ui.adsabs.harvard.edu/abs/2010ApJ...717..825D} {717, 825}

\bibitem[\protect\citeauthoryear{{Everett} \& {Murray}}{{Everett} \&
  {Murray}}{2007}]{2007ApJ...656...93E}
{Everett} J.~E.,  {Murray} N.,  2007, \mn@doi [\apj] {10.1086/510324}, \href
  {https://ui.adsabs.harvard.edu/abs/2007ApJ...656...93E} {656, 93}

\bibitem[\protect\citeauthoryear{{Fichtner}}{{Fichtner}}{1997}]{Fichtner1997}
{Fichtner} H.,  1997, in International Cosmic Ray Conference. p.~397

\bibitem[\protect\citeauthoryear{{Finkbeiner}}{{Finkbeiner}}{2004}]{2004ApJ...614..186F}
{Finkbeiner} D.~P.,  2004, \mn@doi [\apj] {10.1086/423482}, \href
  {https://ui.adsabs.harvard.edu/abs/2004ApJ...614..186F} {614, 186}

\bibitem[\protect\citeauthoryear{{Flynn}, {Sommer-Larsen}  \&
  {Christensen}}{{Flynn} et~al.}{1996}]{1996}
{Flynn} C.,  {Sommer-Larsen} J.,   {Christensen} P.~R.,  1996, \mn@doi [\mnras]
  {10.1093/mnras/281.3.1027}, \href
  {https://ui.adsabs.harvard.edu/abs/1996MNRAS.281.1027F} {281, 1027}

\bibitem[\protect\citeauthoryear{{Fox} et~al.,}{{Fox} et~al.}{2015}]{Fox_2015}
{Fox} A.~J.,  et~al., 2015, \mn@doi [\apjl] {10.1088/2041-8205/799/1/L7}, \href
  {https://ui.adsabs.harvard.edu/abs/2015ApJ...799L...7F} {799, L7}

\bibitem[\protect\citeauthoryear{{Gabici}, {Aharonian}  \& {Blasi}}{{Gabici}
  et~al.}{2007}]{2007Ap&SS.309..365G}
{Gabici} S.,  {Aharonian} F.~A.,   {Blasi} P.,  2007, \mn@doi [\apss]
  {10.1007/s10509-007-9427-6}, \href
  {https://ui.adsabs.harvard.edu/abs/2007Ap&SS.309..365G} {309, 365}

\bibitem[\protect\citeauthoryear{{Gaia Collaboration} et~al.,}{{Gaia
  Collaboration} et~al.}{2016a}]{2016A&A...595A...1G}
{Gaia Collaboration} et~al., 2016a, \mn@doi [\aap]
  {10.1051/0004-6361/201629272}, \href
  {https://ui.adsabs.harvard.edu/abs/2016A&A...595A...1G} {595, A1}

\bibitem[\protect\citeauthoryear{{Gaia Collaboration} et~al.,}{{Gaia
  Collaboration} et~al.}{2016b}]{2016A&A...595A...2G}
{Gaia Collaboration} et~al., 2016b, \mn@doi [\aap]
  {10.1051/0004-6361/201629512}, \href
  {https://ui.adsabs.harvard.edu/abs/2016A&A...595A...2G} {595, A2}

\bibitem[\protect\citeauthoryear{{Gaia Collaboration} et~al.,}{{Gaia
  Collaboration} et~al.}{2018}]{2018A&A...616A...1G}
{Gaia Collaboration} et~al., 2018, \mn@doi [\aap]
  {10.1051/0004-6361/201833051}, \href
  {https://ui.adsabs.harvard.edu/abs/2018A&A...616A...1G} {616, A1}

\bibitem[\protect\citeauthoryear{{Gillessen} et~al.,}{{Gillessen}
  et~al.}{2017}]{2017ApJ...837...30G}
{Gillessen} S.,  et~al., 2017, \mn@doi [\apj] {10.3847/1538-4357/aa5c41}, \href
  {https://ui.adsabs.harvard.edu/abs/2017ApJ...837...30G} {837, 30}

\bibitem[\protect\citeauthoryear{{Guo}}{{Guo}}{2017}]{2017IAUS..322..189G}
{Guo} F.,  2017, {The AGN Jet Model of the Fermi Bubbles}.
pp 189--192 (\mn@eprint {arXiv} {1609.07705}),
  \mn@doi{10.1017/S1743921316012023}

\bibitem[\protect\citeauthoryear{{Guo} \& {Mathews}}{{Guo} \&
  {Mathews}}{2012}]{2012ApJ...756..181G}
{Guo} F.,  {Mathews} W.~G.,  2012, \mn@doi [\apj]
  {10.1088/0004-637X/756/2/181}, \href
  {https://ui.adsabs.harvard.edu/abs/2012ApJ...756..181G} {756, 181}

\bibitem[\protect\citeauthoryear{{Gupta}, {Mathur}  \& {Krongold}}{{Gupta}
  et~al.}{2017}]{2017ApJ...836..243G}
{Gupta} A.,  {Mathur} S.,   {Krongold} Y.,  2017, \mn@doi [\apj]
  {10.3847/1538-4357/836/2/243}, \href
  {https://ui.adsabs.harvard.edu/abs/2017ApJ...836..243G} {836, 243}

\bibitem[\protect\citeauthoryear{{Holzer} \& {Axford}}{{Holzer} \&
  {Axford}}{1970}]{1970ARA&A...8...31H}
{Holzer} T.~E.,  {Axford} W.~I.,  1970, \mn@doi [\araa]
  {10.1146/annurev.aa.08.090170.000335}, \href
  {https://ui.adsabs.harvard.edu/abs/1970ARA&A...8...31H} {8, 31}

\bibitem[\protect\citeauthoryear{{Igarashi}, {Mori}  \& {Nitta}}{{Igarashi}
  et~al.}{2014}]{2014MNRAS.444.1177I}
{Igarashi} A.,  {Mori} M.,   {Nitta} S.-y.,  2014, \mn@doi [\mnras]
  {10.1093/mnras/stu1466}, \href
  {https://ui.adsabs.harvard.edu/abs/2014MNRAS.444.1177I} {444, 1177}

\bibitem[\protect\citeauthoryear{{Igarashi}, {Mori}  \& {Nitta}}{{Igarashi}
  et~al.}{2017}]{2017MNRAS.470.2225I}
{Igarashi} A.,  {Mori} M.,   {Nitta} S.-y.,  2017, \mn@doi [\mnras]
  {10.1093/mnras/stx1349}, \href
  {https://ui.adsabs.harvard.edu/abs/2017MNRAS.470.2225I} {470, 2225}

\bibitem[\protect\citeauthoryear{{Jokipii}}{{Jokipii}}{1966}]{Jokipii_1996}
{Jokipii} J.~R.,  1966, \mn@doi [\apj] {10.1086/148912}, \href
  {https://ui.adsabs.harvard.edu/abs/1966ApJ...146..480J} {146, 480}

\bibitem[\protect\citeauthoryear{{Karim} et~al.,}{{Karim}
  et~al.}{2018}]{2018ApJ...860...98K}
{Karim} T.,  et~al., 2018, \mn@doi [\apj] {10.3847/1538-4357/aac167}, \href
  {https://ui.adsabs.harvard.edu/abs/2018ApJ...860...98K} {860, 98}

\bibitem[\protect\citeauthoryear{Lamers \& Cassinelli}{Lamers \&
  Cassinelli}{1999}]{lamers_cassinelli_1999}
Lamers H. J. G. L.~M.,  Cassinelli J.~P.,  1999, Basic concepts: isothermal
  winds.
Cambridge University Press, p. 60–91, \mn@doi{10.1017/CBO9781139175012.004}

\bibitem[\protect\citeauthoryear{{Lockman}}{{Lockman}}{1984}]{1984ApJ...283...90L}
{Lockman} F.~J.,  1984, \mn@doi [\apj] {10.1086/162277}, \href
  {https://ui.adsabs.harvard.edu/abs/1984ApJ...283...90L} {283, 90}

\bibitem[\protect\citeauthoryear{{Lockman}, {Di Teodoro}  \&
  {McClure-Griffiths}}{{Lockman} et~al.}{2020}]{2020ApJ...888...51L}
{Lockman} F.~J.,  {Di Teodoro} E.~M.,   {McClure-Griffiths} N.~M.,  2020,
  \mn@doi [\apj] {10.3847/1538-4357/ab55d8}, \href
  {https://ui.adsabs.harvard.edu/abs/2020ApJ...888...51L} {888, 51}

\bibitem[\protect\citeauthoryear{{Lynds} \& {Sandage}}{{Lynds} \&
  {Sandage}}{1963}]{1963ApJ...137.1005L}
{Lynds} C.~R.,  {Sandage} A.~R.,  1963, \mn@doi [\apj] {10.1086/147579}, \href
  {https://ui.adsabs.harvard.edu/abs/1963ApJ...137.1005L} {137, 1005}

\bibitem[\protect\citeauthoryear{{Martynenko}}{{Martynenko}}{2022}]{Martynenko_2022}
{Martynenko} N.,  2022, \mn@doi [\mnras] {10.1093/mnras/stac164}, \href
  {https://ui.adsabs.harvard.edu/abs/2022MNRAS.511..843M} {511, 843}

\bibitem[\protect\citeauthoryear{{Mathews} \& {Baker}}{{Mathews} \&
  {Baker}}{1971}]{1971ApJ...170..241M}
{Mathews} W.~G.,  {Baker} J.~C.,  1971, \mn@doi [\apj] {10.1086/151208}, \href
  {https://ui.adsabs.harvard.edu/abs/1971ApJ...170..241M} {170, 241}

\bibitem[\protect\citeauthoryear{{Mertsch} \& {Petrosian}}{{Mertsch} \&
  {Petrosian}}{2019}]{2019A&A...622A.203M}
{Mertsch} P.,  {Petrosian} V.,  2019, \mn@doi [\aap]
  {10.1051/0004-6361/201833999}, \href
  {https://ui.adsabs.harvard.edu/abs/2019A&A...622A.203M} {622, A203}

\bibitem[\protect\citeauthoryear{Mignone, Bodo, Massaglia, Matsakos, Tesileanu,
  Zanni  \& Ferrari}{Mignone et~al.}{2007}]{2007}
Mignone A.,  Bodo G.,  Massaglia S.,  Matsakos T.,  Tesileanu O.,  Zanni C.,
  Ferrari A.,  2007, \mn@doi [\apj] {10.1086/513316}, \href
  {https://ui.adsabs.harvard.edu/abs/2007jena.confR..96M} {170, 228–242}

\bibitem[\protect\citeauthoryear{{Miyamoto} \& {Nagai}}{{Miyamoto} \&
  {Nagai}}{1975}]{1975PASJ...27..533M}
{Miyamoto} M.,  {Nagai} R.,  1975, \pasj, \href
  {https://ui.adsabs.harvard.edu/abs/1975PASJ...27..533M} {27, 533}

\bibitem[\protect\citeauthoryear{{Morris} \& {Serabyn}}{{Morris} \&
  {Serabyn}}{1996}]{1996ARA&A..34..645M}
{Morris} M.,  {Serabyn} E.,  1996, \mn@doi [\araa]
  {10.1146/annurev.astro.34.1.645}, \href
  {https://ui.adsabs.harvard.edu/abs/1996ARA&A..34..645M} {34, 645}

\bibitem[\protect\citeauthoryear{{Mou}, {Yuan}, {Bu}, {Sun}  \& {Su}}{{Mou}
  et~al.}{2014}]{2014ApJ...790..109M}
{Mou} G.,  {Yuan} F.,  {Bu} D.,  {Sun} M.,   {Su} M.,  2014, \mn@doi [\apj]
  {10.1088/0004-637X/790/2/109}, \href
  {https://ui.adsabs.harvard.edu/abs/2014ApJ...790..109M} {790, 109}

\bibitem[\protect\citeauthoryear{{Mou}, {Yuan}, {Gan}  \& {Sun}}{{Mou}
  et~al.}{2015}]{2015ApJ...811...37M}
{Mou} G.,  {Yuan} F.,  {Gan} Z.,   {Sun} M.,  2015, \mn@doi [\apj]
  {10.1088/0004-637X/811/1/37}, \href
  {https://ui.adsabs.harvard.edu/abs/2015ApJ...811...37M} {811, 37}

\bibitem[\protect\citeauthoryear{{Navarro}, {Frenk}  \& {White}}{{Navarro}
  et~al.}{1996}]{1996_NFW}
{Navarro} J.~F.,  {Frenk} C.~S.,   {White} S. D.~M.,  1996, \mn@doi [\apj]
  {10.1086/177173}, \href
  {https://ui.adsabs.harvard.edu/abs/1996ApJ...462..563N} {462, 563}

\bibitem[\protect\citeauthoryear{{Parker}}{{Parker}}{1958}]{1958ApJ...128..664P}
{Parker} E.~N.,  1958, \mn@doi [\apj] {10.1086/146579}, \href
  {https://ui.adsabs.harvard.edu/abs/1958ApJ...128..664P} {128, 664}

\bibitem[\protect\citeauthoryear{{Parker}}{{Parker}}{1965}]{1965ApJ...141.1463P}
{Parker} E.~N.,  1965, \mn@doi [\apj] {10.1086/148235}, \href
  {https://ui.adsabs.harvard.edu/abs/1965ApJ...141.1463P} {141, 1463}

\bibitem[\protect\citeauthoryear{{Particle Data Group} et~al.,}{{Particle Data
  Group} et~al.}{2020}]{2020PTEP.2020h3C01P}
{Particle Data Group} et~al., 2020, \mn@doi [Progress of Theoretical and
  Experimental Physics] {10.1093/ptep/ptaa104}, \href
  {https://ui.adsabs.harvard.edu/abs/2020PTEP.2020h3C01P} {2020, 083C01}

\bibitem[\protect\citeauthoryear{{Planck Collaboration} et~al.,}{{Planck
  Collaboration} et~al.}{2013}]{2013A&A...554A.139P}
{Planck Collaboration} et~al., 2013, \mn@doi [\aap]
  {10.1051/0004-6361/201220271}, \href
  {https://ui.adsabs.harvard.edu/abs/2013A&A...554A.139P} {554, A139}

\bibitem[\protect\citeauthoryear{{Predehl} et~al.,}{{Predehl}
  et~al.}{2020}]{2020}
{Predehl} P.,  et~al., 2020, \mn@doi [\nat] {10.1038/s41586-020-2979-0}, \href
  {https://ui.adsabs.harvard.edu/abs/2020Natur.588..227P} {588, 227}

\bibitem[\protect\citeauthoryear{{Rodgers-Lee}, {Taylor}, {Ray}  \&
  {Downes}}{{Rodgers-Lee} et~al.}{2017}]{2017MNRAS.472...26R}
{Rodgers-Lee} D.,  {Taylor} A.~M.,  {Ray} T.~P.,   {Downes} T.~P.,  2017,
  \mn@doi [\mnras] {10.1093/mnras/stx1889}, \href
  {https://ui.adsabs.harvard.edu/abs/2017MNRAS.472...26R} {472, 26}

\bibitem[\protect\citeauthoryear{{Rodgers-Lee}, {Vidotto}, {Taylor}, {Rimmer}
  \& {Downes}}{{Rodgers-Lee} et~al.}{2020}]{2020MNRAS.499.2124R}
{Rodgers-Lee} D.,  {Vidotto} A.~A.,  {Taylor} A.~M.,  {Rimmer} P.~B.,
  {Downes} T.~P.,  2020, \mn@doi [\mnras] {10.1093/mnras/staa2737}, \href
  {https://ui.adsabs.harvard.edu/abs/2020MNRAS.499.2124R} {499, 2124}

\bibitem[\protect\citeauthoryear{{Sarkar}}{{Sarkar}}{2019}]{2019MNRAS.482.4813S}
{Sarkar} K.~C.,  2019, \mn@doi [\mnras] {10.1093/mnras/sty2944}, \href
  {https://ui.adsabs.harvard.edu/abs/2019MNRAS.482.4813S} {482, 4813}

\bibitem[\protect\citeauthoryear{{Sarkar}, {Nath}  \& {Sharma}}{{Sarkar}
  et~al.}{2017}]{2017MNRAS.467.3544S}
{Sarkar} K.~C.,  {Nath} B.~B.,   {Sharma} P.,  2017, \mn@doi [\mnras]
  {10.1093/mnras/stx314}, \href
  {https://ui.adsabs.harvard.edu/abs/2017MNRAS.467.3544S} {467, 3544}

\bibitem[\protect\citeauthoryear{{Schlickeiser}}{{Schlickeiser}}{1989}]{Schlickeiser_1989}
{Schlickeiser} R.,  1989, \mn@doi [\apj] {10.1086/167009}, \href
  {https://ui.adsabs.harvard.edu/abs/1989ApJ...336..243S} {336, 243}

\bibitem[\protect\citeauthoryear{{Sofue}}{{Sofue}}{2013}]{2013PASJ...65..118S}
{Sofue} Y.,  2013, \mn@doi [\pasj] {10.1093/pasj/65.6.118}, \href
  {https://ui.adsabs.harvard.edu/abs/2013PASJ...65..118S} {65, 118}

\bibitem[\protect\citeauthoryear{{Sofue}}{{Sofue}}{2022}]{2022PASJ..tmp...58S}
{Sofue} Y.,  2022, \mn@doi [\pasj] {10.1093/pasj/psac034}, \href
  {https://ui.adsabs.harvard.edu/abs/2022PASJ..tmp...58S} {}

\bibitem[\protect\citeauthoryear{{Songaila}}{{Songaila}}{1997}]{1997ApJ...490L...1S}
{Songaila} A.,  1997, \mn@doi [\apjl] {10.1086/311006}, \href
  {https://ui.adsabs.harvard.edu/abs/1997ApJ...490L...1S} {490, L1}

\bibitem[\protect\citeauthoryear{{Su}, {Slatyer}  \& {Finkbeiner}}{{Su}
  et~al.}{2010}]{2010}
{Su} M.,  {Slatyer} T.~R.,   {Finkbeiner} D.~P.,  2010, \mn@doi [\apj]
  {10.1088/0004-637X/724/2/1044}, \href
  {https://ui.adsabs.harvard.edu/abs/2010ApJ...724.1044S} {724, 1044}

\bibitem[\protect\citeauthoryear{{Taylor} \& {Giacinti}}{{Taylor} \&
  {Giacinti}}{2017}]{2017PhRvD..95b3001T}
{Taylor} A.~M.,  {Giacinti} G.,  2017, \mn@doi [\prd]
  {10.1103/PhysRevD.95.023001}, \href
  {https://ui.adsabs.harvard.edu/abs/2017PhRvD..95b3001T} {95, 023001}

\bibitem[\protect\citeauthoryear{{Taylor}, {Boylan-Kolchin}, {Torrey},
  {Vogelsberger}  \& {Hernquist}}{{Taylor} et~al.}{2016}]{2016MNRAS.461.3483T}
{Taylor} C.,  {Boylan-Kolchin} M.,  {Torrey} P.,  {Vogelsberger} M.,
  {Hernquist} L.,  2016, \mn@doi [\mnras] {10.1093/mnras/stw1522}, \href
  {https://ui.adsabs.harvard.edu/abs/2016MNRAS.461.3483T} {461, 3483}

\bibitem[\protect\citeauthoryear{{Tsuchiya}, {Mori}  \& {Nitta}}{{Tsuchiya}
  et~al.}{2013}]{2013MNRAS.432.2837T}
{Tsuchiya} M.,  {Mori} M.,   {Nitta} S.-y.,  2013, \mn@doi [\mnras]
  {10.1093/mnras/stt638}, \href
  {https://ui.adsabs.harvard.edu/abs/2013MNRAS.432.2837T} {432, 2837}

\bibitem[\protect\citeauthoryear{{Tumlinson}, {Peeples}  \& {Werk}}{{Tumlinson}
  et~al.}{2017}]{2017ARA&A..55..389T}
{Tumlinson} J.,  {Peeples} M.~S.,   {Werk} J.~K.,  2017, \mn@doi [\araa]
  {10.1146/annurev-astro-091916-055240}, \href
  {https://ui.adsabs.harvard.edu/abs/2017ARA&A..55..389T} {55, 389}

\bibitem[\protect\citeauthoryear{{Vladimirov} et~al.,}{{Vladimirov}
  et~al.}{2011}]{Galprop}
{Vladimirov} A.~E.,  et~al., 2011, \mn@doi [Computer Physics Communications]
  {10.1016/j.cpc.2011.01.017}, \href
  {https://ui.adsabs.harvard.edu/abs/2011CoPhC.182.1156V} {182, 1156}

\bibitem[\protect\citeauthoryear{{Watkins}, {van der Marel}, {Sohn}  \&
  {Evans}}{{Watkins} et~al.}{2019}]{2019ApJ...873..118W}
{Watkins} L.~L.,  {van der Marel} R.~P.,  {Sohn} S.~T.,   {Evans} N.~W.,  2019,
  \mn@doi [\apj] {10.3847/1538-4357/ab089f}, \href
  {https://ui.adsabs.harvard.edu/abs/2019ApJ...873..118W} {873, 118}

\bibitem[\protect\citeauthoryear{{Yang} \& {Ruszkowski}}{{Yang} \&
  {Ruszkowski}}{2017}]{2017ApJ...850....2Y}
{Yang} H. Y.~K.,  {Ruszkowski} M.,  2017, \mn@doi [\apj]
  {10.3847/1538-4357/aa9434}, \href
  {https://ui.adsabs.harvard.edu/abs/2017ApJ...850....2Y} {850, 2}

\bibitem[\protect\citeauthoryear{{Yang}, {Ruszkowski}, {Ricker}, {Zweibel}  \&
  {Lee}}{{Yang} et~al.}{2012}]{2012ApJ...761..185Y}
{Yang} H. Y.~K.,  {Ruszkowski} M.,  {Ricker} P.~M.,  {Zweibel} E.,   {Lee} D.,
  2012, \mn@doi [\apj] {10.1088/0004-637X/761/2/185}, \href
  {https://ui.adsabs.harvard.edu/abs/2012ApJ...761..185Y} {761, 185}

\bibitem[\protect\citeauthoryear{{Yang}, {Ruszkowski}  \& {Zweibel}}{{Yang}
  et~al.}{2013}]{2013MNRAS.436.2734Y}
{Yang} H. Y.~K.,  {Ruszkowski} M.,   {Zweibel} E.,  2013, \mn@doi [\mnras]
  {10.1093/mnras/stt1772}, \href
  {https://ui.adsabs.harvard.edu/abs/2013MNRAS.436.2734Y} {436, 2734}

\bibitem[\protect\citeauthoryear{{Yang}, {Ruszkowski}  \& {Zweibel}}{{Yang}
  et~al.}{2018}]{2018}
{Yang} H.~Y.,  {Ruszkowski} M.,   {Zweibel} E.,  2018, \mn@doi [Galaxies]
  {10.3390/galaxies6010029}, \href
  {https://ui.adsabs.harvard.edu/abs/2018Galax...6...29Y} {6, 29}

\bibitem[\protect\citeauthoryear{{Zhang} \& {Guo}}{{Zhang} \&
  {Guo}}{2021}]{Zhang_2021}
{Zhang} R.,  {Guo} F.,  2021, \mn@doi [\apj] {10.3847/1538-4357/abfdb1}, \href
  {https://ui.adsabs.harvard.edu/abs/2021ApJ...915...85Z} {915, 85}

\bibitem[\protect\citeauthoryear{{de Avillez} \& {Breitschwerdt}}{{de Avillez}
  \& {Breitschwerdt}}{2004}]{Avillez2004}
{de Avillez} M.,  {Breitschwerdt} D.,  2004, \mn@doi [\apss]
  {10.1023/B:ASTR.0000045019.24124.91}, \href
  {https://ui.adsabs.harvard.edu/abs/2004Ap&SS.292..207D} {292, 207}

\makeatother
\end{thebibliography}

\appendix
\section{Hydrostatic density distribution and influence of the outflow}
\label{Hydrostatic}

\begin{figure}
    \centering
    \includegraphics[width=
\linewidth]{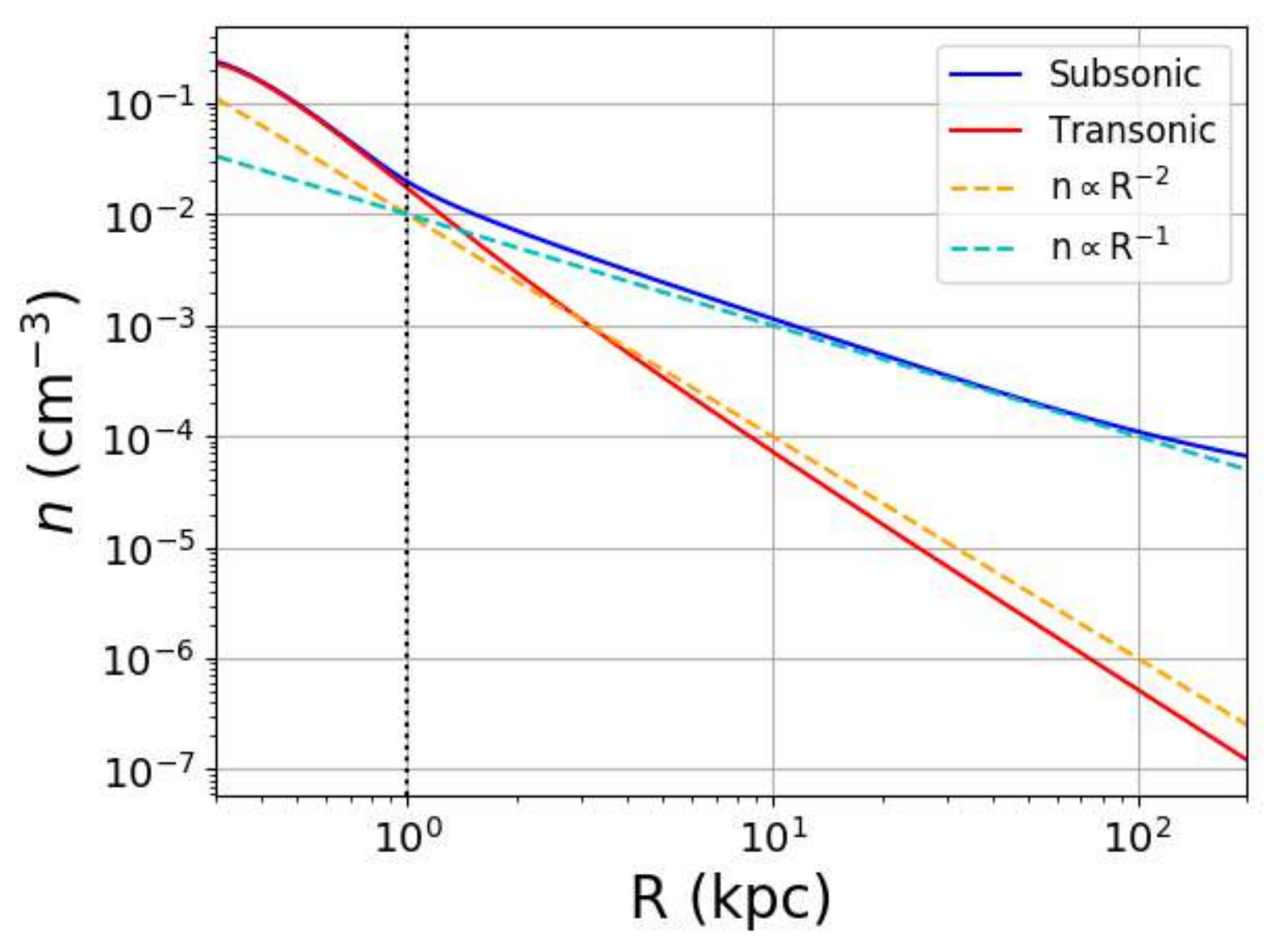}
    \caption{Two density distributions for two specific solutions for the outflow, namely subsonic and transonic. A subsonic solution (solid blue line) leads to a rather flat density distribution, similar to the initial hydrostatic density distribution. This distribution can be approximated  by a power-law distribution,  $n \sim R^{-1}$ (cyan dashed line). A transonic solution (solid red line) produces a steeply falling density distribution beyond the critical radius (black vertical dotted line). This distribution can be approximated by the following power-law distribution, $n \sim R^{-2}$ (orange dashed line).}
    \label{hydrostatic_density}
\end{figure}

Fig.~\ref{hydrostatic_density} shows the density distributions for the subsonic and transonic outflow cases. For the subsonic outflow (blue line), the density distribution is approximately the same as that for the hydrostatic case (defined by Eq.~\ref{density}), whereas the density distribution is changed significantly under the influence of a transonic outflow (red line). It is noted that the initial hydrostatic density distribution is compatible, in the 10-100~kpc distance range, with the constraints from the ram pressure stripping of satellite galaxies around the Milky Way \citep{Martynenko_2022}.

\section{Comparison of the Galactic breeze with observations}
\label{Appendix_B}

\begin{figure}
\includegraphics[width=\linewidth]{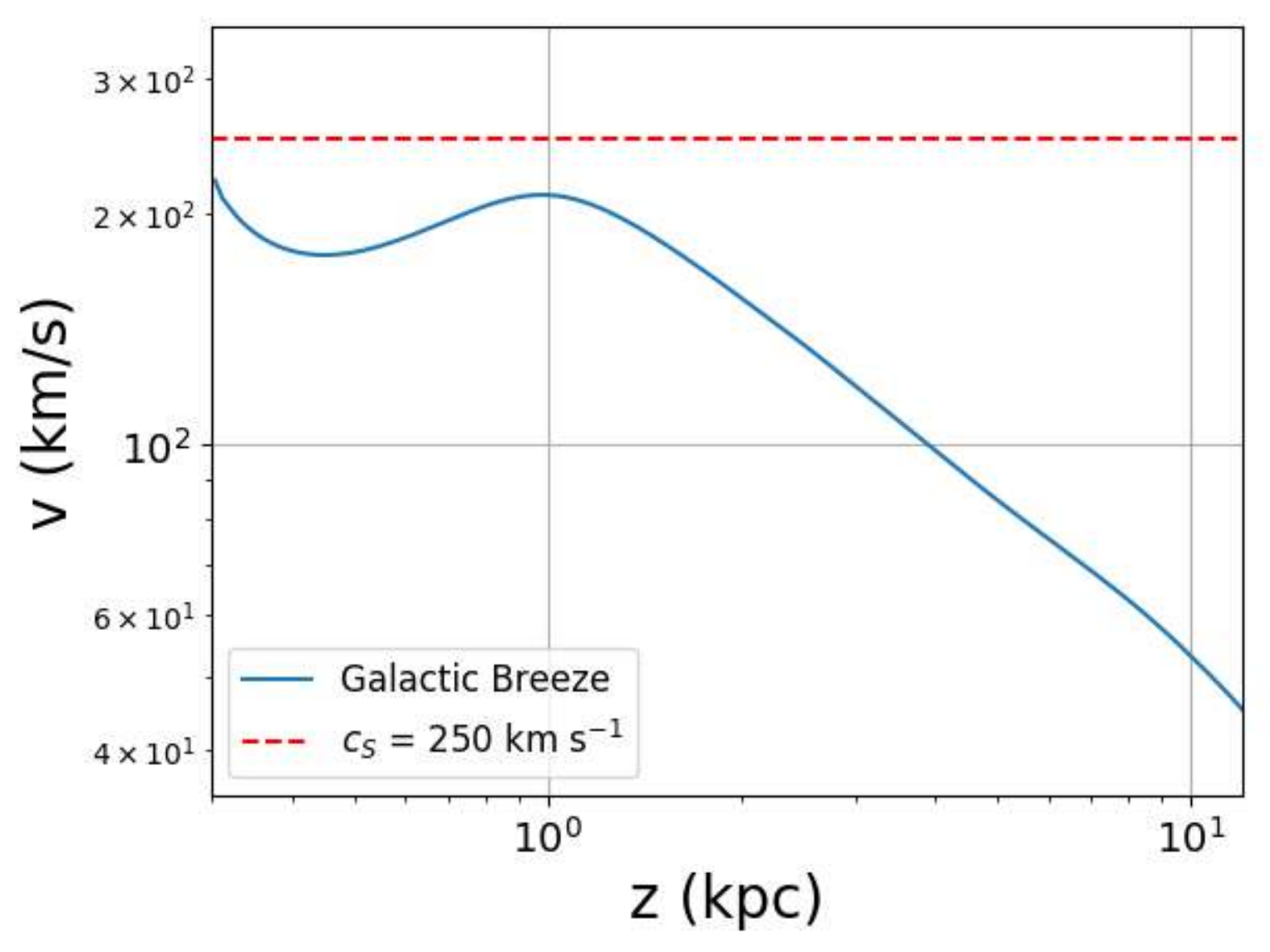}
\caption{The velocity distribution produced by a Galactic breeze (solid blue line)is plotted along the $z$-axis (at $R=0.3$~kpc). The gravitational potential used in this work dictates the behaviour of the outflow. It reaches a maximum at the critical radius (1~kpc) and then decelerates continuously. The dashed red line corresponds to the thermal velocity, $c_s$, for the isothermal gas considered in this paper.}
\label{Velocity_comparison}
\end{figure}

In Fig. \ref{Velocity_comparison} the Galactic breeze produced by the hydrodynamical simulation is presented. As explained in section \ref{Hydrodynamic outflow}, the velocity values for the simulated outflow are too low when compared with observations. Around 1~kpc, the Galactic breeze has a velocity of $\sim$~215~km~s$^{-1}$ where the observations seem to suggest $\sim$~300~km\,s$^{-1}$. However the deceleration of a subsonic outflow profile is still present and seems to match well with the deceleration observed in the North and South Fermi lobes \citep{2017ApJ...834..191B, 2020ApJ...898..128A}.

%\printbibliography
\end{document}